\documentclass{aa}  

\usepackage{graphicx}
\usepackage{txfonts}
\usepackage{xcolor}
\usepackage{subfig}
\usepackage{makecell}
\usepackage{color}
\usepackage[utf8]{inputenc}
\usepackage{multirow}
\usepackage{appendix}
\newcommand{\msun} {\mathrm{M}_\odot}

\newcommand{\vsini} {v\sin\,i}
\newcommand{\kms} {\mathrm{kms}^{-1}}
\newcommand{\denscgs} {\mathrm{gcm}^{-3}}
\newcommand{\ms} {\mathrm{ms}^{-1}}
\newcommand{\days} {\mathrm{d}}
\newcommand{\hours} {\mathrm{h}}
\newcommand{\yr} {\mathrm{yr}}
\newcommand{\myr} {\mathrm{Myr}}
\newcommand{\gyr} {\mathrm{Gyr}}

\newcommand{\kepler} {\emph{Kepler}}
\newcommand{\tess} {\emph{TESS}}

\newcommand{\harps} {HARPS}
\newcommand{\hires} {HIRES}
\newcommand{\pfs} {PFS}
\newcommand{\apf} {APF}

\newcommand{\gj} {GJ\,273}
\newcommand{\gjb} {GJ\,273b}
\newcommand{\gjc} {GJ\,273c}
\newcommand{\gjd} {GJ\,273d}
\newcommand{\gje} {GJ\,273e}
\newcommand{\mearth} {\mathrm{M}_\oplus}
\newcommand{\rearth} {\mathrm{R}_\oplus}
\newcommand{\msini} {M\sin\,i}

\newcommand{\eb} {e_{\mathrm{b}}}
\newcommand{\ec} {e_{\mathrm{c}}}
\newcommand{\ed} {e_{\mathrm{d}}}
\newcommand{\ee} {e_{\mathrm{e}}}

\newcommand{\rb} {R_{\mathrm{b}}}
\newcommand{\rc} {R_{\mathrm{c}}}
\newcommand{\rd} {R_{\mathrm{d}}}
\newcommand{\re} {R_{\mathrm{e}}}

\newcommand{\mb} {M_{\mathrm{b}}}
\newcommand{\mc} {M_{\mathrm{c}}}
\newcommand{\md} {M_{\mathrm{d}}}
\newcommand{\me} {M_{\mathrm{e}}}

\newcommand{\reb} {{\sc rebound}}
\newcommand{\merct} {{\sc mercury-t}}
\newcommand{\merc} {{\sc mercury}}
\newcommand{\possi} {{\sc possidonius}}
\newcommand{\tls} {{\sc tls}}
\newcommand{\atlas} {{\sc atlas2bgeneral}}
\newcommand{\wotan} {{\sc wotan}}
\newcommand{\lk} {{\sc lightkurve}}
\newcommand{\sh} {{\fontfamily{pcr}\selectfont  SHERLOCK}}
\newcommand{\tlsq} {{\sc Transit Least Square}}
\newcommand{\parsec} {{\sc parsec}}

\newcommand{\au} {\mathrm{au}}

\newcommand\isotope[2]{\textsuperscript{#2}#1}

\begin{document}

   \title{GJ~273: On the formation, dynamical evolution and habitability of a planetary system hosted by an M dwarf at 3.75 parsec}
   
   \subtitle{}

   \author{Francisco J. Pozuelos\inst{1,2,$\dag$}
          \and
          Juan C. Su\'arez\inst{3,4}
          \and
          Gonzalo C. de El\'{\i}a\inst{5,6}
          \and
          Zaira M. Berdiñas\inst{7}
          \and
          Andrea Bonfanti\inst{1,10}
          \and
          Agust\'{\i}n Dugaro\inst{5,6}        
          \and
          Michaël Gillon\inst{2}
          \and
          Emmanuël Jehin\inst{1}
          \and
          Maximilian N. Günther\inst{8,9}
          \and
          Val\'erie Van Grootel\inst{1}
          \and
          Lionel J. Garcia\inst{2}
          \and
          Antoine Thuillier\inst{1,11}
          \and
          Laetitia Delrez\inst{1,2}
          \and 
          Jose R. Rod\'on\inst{4}
          }

   \institute{
             Space sciences, Technologies \& Astrophysics Research (STAR) Institute, Universit\'e de Li\`ege, 4000 Li\`ege, Belgium 
             \and EXOTIC Lab, UR Astrobiology, AGO Department, University of Li\`ege, 4000 Li\`ege, Belgium 
             \and Dpt. F\'isica Te\'orica y del Cosmos, Universidad de Granada, Campus de Fuentenueva s/n, 18071, Granada, Spain 
             \and Instituto de Astrof\'isica de Andaluc\'ia (CSIC), Glorieta de la Astronom\'ia s/n, 18008, Granada, Spain
             \and Instituto de Astrof\'isica de La Plata, La Plata , Argentina
             \and Facultad de Ciencias Astron\'omicas y Geof\'isicas, Universidad Nacional de La Plata.  LaPlata, Argentina
             \and Departamento de Astronom\'ia, Universidad de Chile, Camino El Observatorio 1515, Las Condes, Santiago, Chile, Casilla 36-D
             \and Department of Physics, and Kavli Institute for Astrophysics and Space Research, Massachusetts Institute of Technology, Cambridge, MA 02139, USA
             \and Juan Carlos Torres Fellow
             \and Space Research Institute, Austrian Academy of Sciences, Schmiedlstrasse 6, A-8042 Graz, Austria
             \and Observatoire de Paris, 61 Avenue de l’Observatoire 75014 Paris, France \\
             $\dag$~\email{fjpozuelos@uliege.be}
             }

  \date{Received ; accepted }

 
  \abstract
   {Planets orbiting low-mass stars such as M dwarfs are now considered a cornerstone in the search for life-harbouring planets. \gj\ is a planetary system orbiting an M dwarf only 3.75~pc away, composed of two confirmed planets, \gjb\ and \gjc, and two promising candidates, \gjd\ and \gje. Planet \gjb\ resides in the habitable zone. Currently, due to a lack of observed planetary transits, only the minimum masses of the planets are known: M$_{b}$~$\sin i_{b}$=2.89~M${_\oplus}$, M$_{c}$~$\sin i_{c}$=1.18~M${_\oplus}$, M$_{d}$~$\sin i_{d}$=10.80~M${_\oplus}$, and M$_{e}$~$\sin i_{e}$=9.30~M$_{\oplus}$. Despite being an interesting system, the \gj\ planetary system is still poorly studied.}
   {We aim at precisely determine the physical parameters of the individual planets, in particular to break the mass--inclination degeneracy to accurately determine the mass of the planets. Moreover, we present thorough characterisation of planet \gjb\ in terms of its potential habitability.}
  {First, we explored the planetary formation and hydration phases of \gj\ during the first $100\,\myr$. Secondly, we analysed the stability of the system by considering both the two- and four-planet configurations. We then performed a comparative analysis between \gj\ and the Solar System, and searched for regions in \gj\ which may harbour minor bodies in stable orbits, i.e. main asteroid belt and  Kuiper belt analogues.}
   {From our set of dynamical studies, we obtain that the four-planet configuration of the system allows us to break the mass--inclination degeneracy. From our modelling results,
   the masses of the planets are unveiled as: $2.89\leq \mb\leq3.03\,\mearth$, $1.18\leq \mc\leq1.24\,\mearth$, $10.80\leq \md\leq11.35\,\mearth$ and $9.30\leq \me\leq9.70\,\mearth$. These results point to a system likely composed of an Earth-mass planet, a super-Earth and two mini-Neptunes. From planetary formation models, we determine that \gjb\ was likely an efficient water captor while \gjc\ is probably a dry planet. We found that the system may have several stable regions where minor bodies might reside. Collectively, these results are used to comprehensively discuss the habitability of \gjb.}
   {}

   \titlerunning{GJ~273; a nearby system}
   \authorrunning{Pozuelos et al. } 
   
   \maketitle

%

\section{Introduction} \label{intro}

In the last decade, interest in planetary systems around M-type stars has significantly grown. The scientific interest is twofold. On one side, M-type stars are the most abundant star type in the Solar neighbourhood, and indeed the Universe \citep[see e.~g.,][]{Henry1994TheParsecs, Chabrier2000TheoryObjects,Winters2015TheSky}. However, their physical properties are poorly known since they are generally quite faint, as compared with other stellar types, and have only recently become the cornerstone in the detection of low-mass planets. Indeed, recent studies based on radial--velocity (RV) and transit photometry surveys have hinted at the high occurrence of low-mass planets orbiting M dwarfs, with a rate of at least one planet per star \citep[e. g.,][]{bonfils2013,dressing2013,dressing2015,tuomi2014}. This fact is also highlighted by the number of dedicated surveys targeting M dwarfs, notably in the search of rocky planets in their habitable zones (HZs). Some examples are the SPECULOOS \citep{Gillon2018SearchingWorlds,Burdanov2018SPECULOOSTRAPPIST,Delrez2018SPECULOOS:Dwarfs,Jehin2018ThePlanets}, CARMENES \citep{Quirrenbach2010CARMENES:Spectrograph, Reiners2018TheStars}, and MEarth \citep{Nutzman2008DesignDwarfs} projects. 

More interestingly, the huge number of M-type stars has significantly increased the chances of finding Earth-mass rocky planets in temperate orbits, i.e. within the host star’s HZ. Nevertheless, planets in the HZs of M dwarfs are close to the star and thus are generally exposed to intense radiation and magnetic fulgurations. Whether a planet can remain habitable in the presence of UV and X-ray flares is still a matter of debate: while a single flare probably does not compromise a planet's habitability, planets orbiting M dwarfs may suffer frequent and strong flares at younger ages \citep{armstrong2016}.

Among all the planets discovered so far orbiting M dwarfs, some of the most interesting are also the closest to Earth. One such example is the temperate Earth-mass planet orbiting Proxima Centauri-b, which is located only 1.2~pc away \citep{Anglada-Escude2016ACentauri}, while the seven Earth-size planets orbiting the TRAPPIST-1 system at a distance of 12.1~pc, where at least three of them are in the HZ \citep{Gillon2016TemperateStar,Gillon2017SevenTRAPPIST-1} are equally tantalizing. In this context, \gj\ is a highly interesting M dwarf (also known as Luyten's star) hosting a multi-planet system, which is one of the closest stars to the Sun located at only 3.75~pc \citep{Gatewood2008AstrometricRegions}. 

Compared with other M-type stars, \gj\ is a bright star \citep[9.87 Vmag, ][]{koen2010}, with a main-sequence spectral type of M3.5V \citep{Hawley1996TheActivity}, which shows a small projected rotational velocity \citep[$\vsini< 2.5\,\kms$, ][]{browning2010}. In the work by \cite{astudillo2017}, the authors made use of a data set of 280 RVs measured with \harps\ (mounted on ESO's 3.5-m La Silla telescope in Chile) which covered 12 years of intense monitoring of \gj. \cite{astudillo2017} reported the presence of two short-period planets, \gjb\ and \gjc, with minimum masses of 2.89 and 1.18~M$_{\oplus}$ and orbital periods of 18.6 and 4.7~d, respectively. Planet \gjb\ was shown to be located in the HZ of \gj, which along with its close proximity to Earth, sparked interest in this particular system. In addition, in the recent study presented by \cite{Tuomi2019FrequencyNeighbourhood} the authors made use of a data set that combined \harps, \hires\ (Keck), \pfs\ (Subaru) and \apf\ (Lick), for a total of 466 RVs measurements, and fully confirmed the existence of the two planets previously detected by \cite{astudillo2017} and suggested the presence of two extra planets at larger heliocentric distances: \gjd\ with a minimum mass of 10.8~M$_{\oplus}$ and an orbital period of 413.9~d, and \gje\ with a minimum mass of 9.3~M$_{\oplus}$ and an orbital period of 542~d\footnote{At the time of writing, the study of \cite{Tuomi2019FrequencyNeighbourhood} is still being peer-reviewed.}. The authors claimed these new signals as \emph{planetary candidates} which need more data to verify their planet nature. In addition,  \cite{astudillo2017} reported the presence of long-period signals with P$\sim$420 and $\sim$700~d. However, both signals were not considered by the authors as planetary candidates, but were due instead to stellar activity. Previously to that, \cite{bonfils2013} also found a controversial signal of a possible planet at P$\sim$420~d, however, due to their poor phase coverage, the authors considered it an unclear detection. 

Collectively, the present data have prevented us from making any definite statements regarding the precise number of planets in the \gj\ system. Therefore, since the existence of the two outermost planets \gjd\ and \gje\ is still under debate, in the present study we consider two options, two-- and four--planet configurations. In Table~\ref{table:sys_prop}, we have presented the known physical the characteristics of the planets and the star.         

The paper is organized as follows: in Section~2, we attempt to reproduce the system architecture using accretion simulations where we track the content of water of the planets. In 
Section~3, we review the current observational data set that consists of RVs and photometric measurements. In Section~4, we analyse the stability of the system and provide extra constraints on the planetary parameters and the general system architecture, while in Section~5 we explore the similarities of \gj\ with the Solar System while searching for minor-body reservoir analogues. In Section~6, we discuss on the habitability of planet \gjb, and finally, in Section~7 we present our conclusions.

\begin{table*}
\caption{\gj\ system's properties}             
\label{table:sys_prop}      
\centering 
\begin{tabular}{l c c c c}     
\hline\hline       
\noalign{\smallskip}                     
\multicolumn{5}{c}{Star}\\
\noalign{\smallskip} 
\hline 
\noalign{\smallskip}
\multirow{3}{*}{Other designations} & \multicolumn{4}{c}{BD+05 1668} \\
 & \multicolumn{4}{c}{2MASS J07272450+0513329} \\ 
 & \multicolumn{4}{c}{DR2 3139847906304421632} \\
Spect. Type$^{(1)}$  & \multicolumn{4}{c}{M3.5V}   \\
$\alpha$ (J2000) & \multicolumn{4}{c}{$07^{h}27^{m}24.49^{s}$}   \\
$\delta$ (J2000) & \multicolumn{4}{c}{$+05\degr13^{\prime}32.8^{\prime\prime}$} \\
$V$ $^{(2)}$ & \multicolumn{4}{c}{9.872} \\
$J$ $^{(3)}$ & \multicolumn{4}{c}{$5.714\pm0.032$} \\
$H$ $^{(3)}$ & \multicolumn{4}{c}{$5.219\pm0.063$} \\
$K_{\mathrm{s}}$ $^{(3)}$& \multicolumn{4}{c}{$4.857\pm0.023$} \\
$L~(\mathrm{L}_{\odot})$ $^{(4)}$ & \multicolumn{4}{c}{0.0088} \\
$R~(\mathrm{R}_{\odot})$ $^{(4)}$ & \multicolumn{4}{c}{$0.293\pm0.027$} \\
$T_{\mathrm{eff}}~(\mathrm{K})$ $^{(4)}$ & \multicolumn{4}{c}{$3382\pm49$} \\
$M~(\mathrm{M}_{\odot})$ $^{(5)}$ & \multicolumn{4}{c}{0.29} \\
$[\mathrm{Fe/H}]$ $^{(6)}$ & \multicolumn{4}{c}{$0.09\pm0.17$} \\
P$_{rot}~(\mathrm{d})$ $^{(7)}$ & \multicolumn{4}{c}{99} \\ 
\noalign{\smallskip} 
\hline 
\hline 
\noalign{\smallskip} 
\multicolumn{5}{c}{Planets}\\
\noalign{\smallskip} 
\hline 
\noalign{\smallskip}
Planet & P~(d) & M~$\sin{i}$~(M$_{\oplus}$)  & a~(au) & e  \\
\hline 
\noalign{\smallskip}
\gjb\ $^{(8)}$ & 18.6498$^{+0.0059}_{-0.0052}$ & 2.89$^{+0.027}_{-0.26}$  & 0.091101$^{+0.000019}_{-0.000017}$ & 0.10$^{+0.09}_{-0.07}$  \\
\noalign{\smallskip}
\gjc\ $^{(8)}$ & 4.7234$^{+0.0004}_{-0.0004}$ & 1.18$^{+0.016}_{-0.16}$  & 0.036467$^{+0.000002}_{-0.000002}$ & 0.17$^{+0.13}_{-0.12}$  \\
\noalign{\smallskip}
\gjd\ $^{(9)}$ & 413.9$^{+4.3}_{-5.5}$ & 10.8$^{+3.9}_{-3.5}$  & 0.712$^{+0.062}_{-0.076}$ & 0.17$^{+0.18}_{-0.17}$  \\
\noalign{\smallskip}
\gje\ $^{(9)}$  & 542$^{+16}_{-16}$ & 9.3$^{+4.3}_{-3.9}$  & 0.849$^{+0.083}_{-0.092}$ & 0.03$^{+0.20}_{-0.03}$  \\
\noalign{\smallskip}
\hline
\noalign{\smallskip}

\end{tabular}
\tablefoot{ (1)~\cite{Hawley1996TheActivity}; (2)~\cite{Gaidos2014TrumpetingLife}; (3)~\cite{Cutri2003VizieR2003}; (4)~\cite{Boyajian2012StellarM-stars}; 
(5)~\cite{Delfosse2000AccurateRelations};
(6)~\cite{Neves2013MetallicitySample};
(7)~\cite{Astudillo-Defru2017MagneticRHK};
(8)~\citet{astudillo2017};  (9)~\cite{Tuomi2019FrequencyNeighbourhood}.  
            }
\end{table*}


\section{Formation and evolution of the planetary system GJ~273: N-body experiments}

In the present section, we analyse the processes of formation and evolution of the planetary system \gj\ using N-body simulations. To do this, we first describe the properties of the protoplanetary disk and then determine the physical and orbital initial conditions to be used in our model. Then, we define the different scenarios to be modelled and finally, we discuss the results obtained from our N-body experiments. 

\subsection{Protoplanetary disk: properties}

An important parameter that determines the distribution of the material in a protoplanetary disk is the surface density. The gas-surface density profile $\Sigma_{\textrm{g}}(R)$ adopted in the present research is given by 
\begin{equation}
\Sigma_{\textrm{g}}(R)=\Sigma_{0\textrm{g}}\left(\frac{R}{R_{\textrm{c}}}\right)^{-\gamma}\textrm{e}^{{-(R/R_{\textrm{c}})}^{2-\gamma}},
\label{eq:densgas}
\end{equation}
where $R$ is the radial coordinate in the disk mid-plane, $\Sigma_{0\textrm{g}}$ a normalization constant, $R_{\textrm{c}}$ the characteristic radius, and $\gamma$ an exponent that determines the density gradient. At the same way, the solid-surface density profile $\Sigma_{\textrm{s}}(R)$ is written by 
\begin{equation}
\Sigma_{\textrm{s}}(R)=\Sigma_{0\textrm{s}}\eta_{\textrm{ice}}\left(\frac{R}{R_{\textrm{c}}}\right)^{-\gamma}\textrm{e}^{{-(R/R_{\textrm{c}})}^{2-\gamma}},
\label{eq:denssol}
\end{equation}
where $\Sigma_{0\textrm{s}}$ is a normalization constant and parameter $\eta_{\textrm{ice}}$ represents the change in the amount of solid material due to the condensation of water beyond the snow line, which is assumed to be initially located at $R_{\textrm{ice}}$.

To determine the normalization constant $\Sigma_{0\textrm{g}}$, we integrated the expression concerning the mass of the protoplanetary disk assuming axial symmetry. From this, we can express the total mass of the disk $M_{\text{d}}$ by
\begin{equation}
  M_{\textrm{d}}=\displaystyle\int_{0}^{\infty} 2\pi R\Sigma_{\textrm{g}}(R)\,\textrm{d}R,
  \label{eq:mdisco}
\end{equation}
from which we obtain
\begin{equation}
\Sigma_{0\textrm{g}} = (2 - \gamma) \frac{M_{\textrm{d}}}{2\pi R_{\textrm{c}}^{2}}.
\end{equation}
The normalization constant $\Sigma_{0\textrm{s}}$ can be obtained making use of the relation
\begin{equation}
  \Sigma_{0\textrm{s}} = z_{0} \Sigma_{0\textrm{g}} 10^{-\text{[Fe/H]}},
  \label{eq:ratioperfiles}
\end{equation}
where [Fe/H] is at stellar metallicity and $z_{0}$ is the primordial abundance of heavy elements in the sun, which has a value of $z_{0}=0.0153$ \citep{Lodders20094.4System}.

To quantify the gas- and solid-surface density profiles involved in our model, we must assign values to several parameters such as the stellar metallicity [Fe/H], the mass of the disk $M_{\textrm{d}}$, the characteristic radius $R_{\textrm{c}}$, the exponent $\gamma$, the location of the snow line $R_{\textrm{ice}}$ and the $\eta_{\textrm{ice}}$ parameter. On the one hand, the \gj\ star has a mass of $M_{\star}$ = 0.29~M$_{\odot}$ and a metallicity of [Fe/H] = 0.09 \citep{astudillo2017}. On the other hand, our model uses a disk mass of $M_{\textrm{d}} = 5~\% \textrm{M}_{\star}$, a characteristic radius of $R_{\textrm{c}} = 20\,\au$, and an exponent of $\gamma =$ 1, which are consistent with observational studies of protoplanetary disks carried out by several authors \citep[e.g.][]{Andrews2010ProtoplanetarySources, Testi2016BrownOphiuchus, Cieza2019TheResolution}, and with population synthesis models of protostellar discs \citep{Bate2018OnDiscs}.
The position of the snow line $R_{\textrm{ice}}$ must be carefully chosen. To do this, we followed the procedure proposed by \citet{Ciesla2015VolatileStars}, who studied the delivery of volatiles to planets from water-rich planetesimals around low-mass stars. That work assumed that $R_{\textrm{ice}}$'s location evolves in a disk as a result of the variations over time of the stellar radius and effective temperature. With such an assumption, the snow line is about at 0.65~au (disk temperature of 140~K, age of 1~Myr) when the stellar parameters given by \citet{Siess2000AnStars} are considered. Although the location of the sow line evolves in time, we followed \citet{Ciesla2015VolatileStars} by assuming a fixed value for $R_{\textrm{ice}}$ for the whole duration of our N-body experiments (see Apendix~\ref{ap:sl}).


It is also important to consider that the location of the snow line generally represents the transition from dry to water-rich bodies in the disk. However, any increase in the amount of solid material due to the condensation of water beyond the snow line is strongly dependent on the existence of short-lived radionuclides, notably \isotope{26}{Al}. According to the work carried out by \citet{Lodders2003SolarElements} about Solar System abundances, we consider that the $\eta_{\textrm{ice}}$ parameter has a value of 1 and 2 inside and beyond $R_{\textrm{ice}}$, respectively. From this, we assume that bodies initially located inside the snow line are primarily made out of rocky material with only 0.01$\%$ water by mass, while the initial content of water by mass of those bodies associated with the outer region of the system beyond the snow line is of order 50$\%$.

Our investigation is focused on the formation, evolution, and survival of the confirmed planets in \gj, specially the one close to the HZ. Thus, it is necessary to specify the limits of such a region in the system under study. 
We derive conservative and optimistic estimates for the width of the HZ around \gj\ from the models developed by \citet{Kopparapu2013HabitableEstimates} of 0.1--$0.19\,\au$ and 0.08--$0.2\,\au$, respectively. We assume that a planet is in the HZ of the system if its whole orbit is contained inside the optimistic edges. According to this, a planet was considered to be in the HZ of the system if its pericentric distance was $q \geq 0.08\,\au$ and its apocentric distance was $Q\leq0.2\,\au$.

From that described in this section, we consider that the extension of the region of study is between $0.01\,\au$ and $1.5\,\au$, since it includes the HZ, the snow line, and an outer water-rich embryo population.

\begin{figure*}
\includegraphics[width=0.85\textwidth]{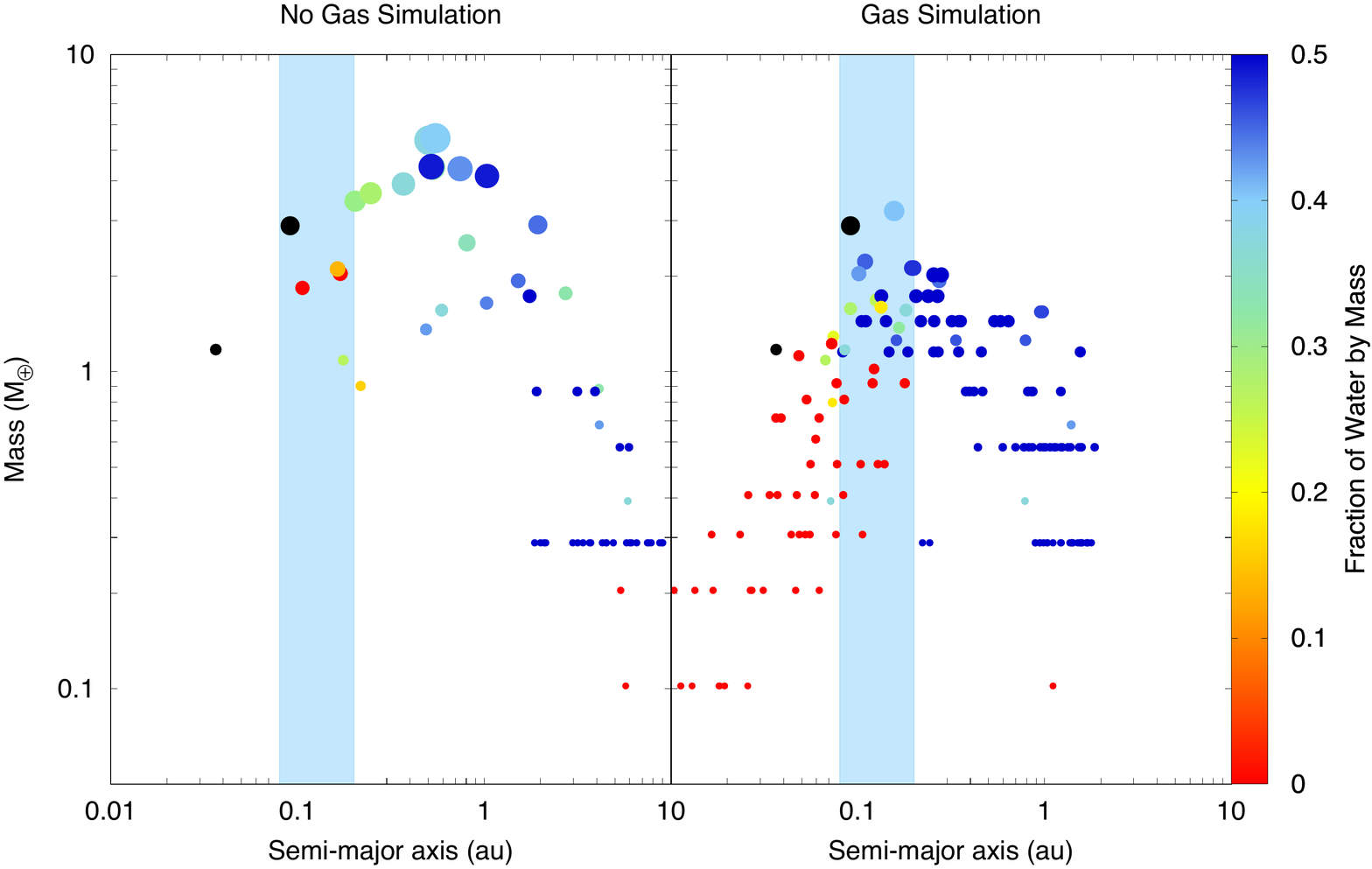}
\caption{Mass distribution of the planets formed in the N-body simulations after $100\,\myr$ of Set 1 (left panel), which only model the post-gas stage, and Set 2 (right panel), which include the effects of the gas disk. In every panel, the black circles illustrate the two planets confirmed that orbit \gj\ \citep{astudillo2017}, while the color code represents the final fraction of water by mass of the planets resulting from our numerical experiments. Finally, the light blue shaded region refers to the optimistic HZ derived from the model developed by \citet{Kopparapu2013HabitableEstimates,Kopparapu2013Erratum:131}.}
\label{fig:figura_masas_vs_a}
\end{figure*}

\subsection{N-body experiments: characterization, parameters, and initial conditions}

In this sub-section, we analyse the process of formation and evolution of terrestrial-like planets in the \gj\ system from N-body experiments. 
The \merc\ N-body code was used to develop our simulations \citep{Chambers1999ABodies}. In particular, we made use of a hybrid integrator, which uses a second-order mixed variable symplectic algorithm to treat the interactions between objects with separations greater than 3~Hill radii, and the Bulirsch--Stoer method for resolving closer encounters.

In general terms, the accretion models were aimed at studying the formation of terrestrial-like planets via N-body experiments to track the evolution of planetary embryos and planetesimals in a given system. However, N-body simulations that include a population of embryos and thousands of planetesimals are very costly numerically. In the present research, we considered a simplified model, which assumes a disk only composed of planetary embryos. The \merc\ code computes the temporal evolution of the orbits of the embryos by simulating the gravitational interactions among them. Such interactions produce orbits that cross each other, which results in successively close encounters between them, which in turn lead to collisions among them, ejections from the system and collisions with the central star. Several works have shown that collisions between gravity-domained bodies yield different outcomes depending on the target size, projectile size, impact velocity, and impact angle \citep{Leinhardt2012CollisionsLaws, Chambers2013Late-stageFragmentation, Quintana2016THEWORLDS, Dugaro2019PhysicalFragmentation}. In the present study, all collisions were treated as perfect mergers, conserving the total mass of the interacting bodies in each impact event. This approach has been the main hypothesis of multiple works concerning accretion simulations aimed at studying the process of planetary formation in systems with different dynamical scenarios \citep[e.g.][]{Raymond2004MakingDelivery, OBrien2006TerrestrialFriction, Lissauer2007PlanetsVolatiles, Raymond2009BuildingSystem, deElia2013TerrestrialGiants, Dugaro2016TerrestrialStars, Darriba2017MigratingDelivery, Zain2018PlanetaryEnvironments}.

Our N-body simulations assumed 53 planetary embryos between $0.01\,\au$ and $1.5\,\au$, 28 (25) of which were initially located before (beyond) the snow line at $0.65\,\au$, and possessed individual masses and physical densities of $m_{\textrm{p}} = 0.102 (0.28)\,\mearth$ and $\rho_{\textrm{p}} = 3\,(1.5)\,\denscgs$, respectively. The embryos were distributed across the region of study using the acceptance--rejection method developed by Jonh von Neumann from the function
\begin{equation}
dN(R) = \frac{2 \pi}{m_{\textrm{p}}} R \Sigma_{\textrm{s}}(R) dR,
\label{eq:numeroemb}
\end{equation}
where $\Sigma_{\textrm{s}}(R)$ is represented by Eq.~\ref{eq:denssol}. Given that the planetary embryos were assumed to start the simulation in quasi-circular and coplanar orbits, the values of the variable $R$ generated from Eq.~\ref{eq:numeroemb} were adopted as their initial semi-major axes $a$. In fact, the initial eccentricities $e$ and inclinations $i$ of the embryos were calculated randomly from uniform distributions assuming values lower than 0.02 and 0.5$^{\circ}$, respectively. In the same way, the argument of the pericenter $\omega$, the longitude of the ascending node $\Omega$ and the mean anomaly $M$ of the planetary embryos were chosen randomly from uniform distributions between 0$^{\circ}$ and 360$^{\circ}$.

The N-body experiments developed using the hybrid integrator of the \merc\ code required a good selection of the time step to carry out a correct integration of the system of study. According to this, we used a time step of 0.03~d, which is shorter than 1/20th of the orbital period of the innermost body in our scenario of work. Moreover, in order to avoid the integration of orbits with very small pericenters, our model assumed a non-realistic value for the radius of the central star of $0.005\,\au$. Finally, we considered an ejection distance of $1000\,\au$ in all our N-body experiments.

In order to analyse the formation and evolution history of the planets that compose system \gj, we carried out two sets of N-body accretion simulations:
\begin{itemize}
\item {\it Set 1,} which analyzes the post-gas phase; 
\item {\it Set 2,} which includes the effects of the gas disk.
\end{itemize}
`Set 2' N-body simulations included modelling of the effects of the gaseous component of the disk in the early evolution of the planetary embryos. To do this, we developed an improved version of the \merc\ code, which included analytical prescriptions for the orbital migration rates and eccentricity and inclination damping rates derived by \citet{Cresswell2008Three-dimensionalDisc} from the formulae for damping rates obtained by \citet{Tanaka2004ThreedimensionalWaves} and the migration rates obtained by \citet{Tanaka2002ThreedimensionalMigration}. We will give a detailed description about this in Section~2.4.  

Due to the stochastic nature of the accretion process, we carried out between 10 and 15 runs for each set of numerical simulations. We must remark that, in both of them, each N-body experiment was integrated for $100\,\myr$.

\subsection{`Set 1': Collisional accretion of protoplanets after gas dissipation}

In this set of N-body experiments, we considered that the initial physical and orbital parameters assigned to the planetary embryos in Section 2.2 were associated with the post-gas phase. Thus, here, our main goal is to study the formation and evolution of terrestrial-like planets via the collisional accretion of embryos once the gas has dissipated from the system. 

The left panel of Fig.~\ref{fig:figura_masas_vs_a} shows the mass distribution as a function of the semi-major axis of the planets formed in 9 N-body simulations of `Set 1' after 100~Myr of integration. 
Moreover, the light blue shaded area illustrates the optimistic limits of the HZ of the system. 
The distribution of planets 
offers two important results. On the one hand, the most massive planets in this scenario formed at locations close to the snow line, which was initially at around $0.65\,\au$. In fact, our N-body experiments produced massive planets of about $5.4\,\mearth$ and semi-major axes ranging between $0.50\,\au$ and $0.55\,\au$. On the other hand, our N-body simulations did not efficiently produce planets in the HZ nor in the inner region of the system. In fact, only three planets formed in the HZ with masses between $1.6\,\mearth$ and $2.1\,\mearth$, while no planet survived in the inner region of the system at $100\,\myr$. According to this, this scenario was not able to produce planets analogous to those observed in the real system around \gj. 

\subsection{`Set 2': Effect of the gas disk}

In this set of N-body experiments, we tested the effects of the presence of a gas disk in the early evolution of the planetary embryos. To do this, we included additional forces to model the effects of a dissipating gaseous disk on the orbital parameters of the planetary embryos. In particular, our improved version of the \merc\ code included effects that led to changes of the semi-major axis (hereafter, `type-1 migration') as well as decays in eccentricity and inclination (hereafter, `type-1 damping') of the embryos. 

The type-1 migration and type-1 damping were modelled based on hydrodynamic simulations focused on planetary embryos embedded within idealized isothermal disk \citep{Tanaka2002ThreedimensionalMigration, Tanaka2004ThreedimensionalWaves}. However, those works are only valid for embryos that are assumed to have no disk gap and with low values associated with their eccentricities and inclinations. In order to develop a more general treatment, we modified the model to include the additional terms derived by \citet{Cresswell2008Three-dimensionalDisc}, which allowed us to describe the type-1 migration and type-1 damping for embryos evolving on orbits with large eccentricities and inclinations.


In idealized vertically isothermal disks, the type-1 migration derived by \citet{Tanaka2002ThreedimensionalMigration} can led to the very fast inward orbital migration of the planetary embryos. In general terms, the migration times calculated with this model were comparable to or smaller than the known lifetimes of gas disks, which may be at odds with
current planetary formation models. Type-1 migration tends to be more complex when a more detailed treatment concerning the physics of the disk is considered \citep[e.g.][]{Paardekooper2010ADrag, Paardekooper2011ADiffusion, Guilet2013TypeDiscs, Benitez-Llambay2015PlanetCores}. Beyond this, several authors have carried out population synthesis studies using the expressions of \citet{Tanaka2002ThreedimensionalMigration} concerning type-1 migration rates, with an added reduction factor in order to reproduce observations \citep[e.g.][]{Alibert2005ModelsEvolution, Mordasini2009ExtrasolarObservations, Miguel2011TheObservations, Ronco2017FormationAnalysis}.    

Here, we considered the type-1 migration rate for isothermal disks of \citet{Tanaka2002ThreedimensionalMigration} and \citet{Cresswell2008Three-dimensionalDisc}, and we also incorporated a factor, $f_1$, that ranged between 0 and 1. According to this, the type-1 migration time is inversely proportional to such a factor, for which the larger the value of $f_1$, the faster the orbital migration. From this, we carried out N-body experiments aimed at studying the gas effects on the formation and evolution of the planetary system under consideration for values of $f_1$ between 0.01 and 1. Naturally, our results showed that the larger the value of $f_1$, the smaller the semi-major axes of the resulting planets.    

The gas disk was modelled by adopting the gas-surface density profile $\Sigma_{\textrm{g}}(R)$ given by Eq.~\ref{eq:densgas} and a height scale of $H = 0.0361 (R/1\textrm{ua})^{5/4}\,\au$. In this scenario, we considered that the gaseous component dissipates exponentially over $2.5\,\myr$, which is consistent with studies concerning the lifetimes of primordial disks associated with young stellar clusters developed by \citet{Mamajek2009InitialDisks}. According to this, it is very important to remark that the zero-time of our N-body experiments corresponds to $1\,myr$, for which the gas effects were modelled for $1.5\,\myr$ in all the numerical simulations.

The right panel of Fig.~\ref{fig:figura_masas_vs_a} illustrates the mass of the planets formed in 14 N-body simulations of `Set 2' after $100\,\myr$ of integration as a function of the semi-major axis, assuming a reduction factor $f_1$ for a type I migration of 0.04. According to this, the most massive planets produced in this scenario were located in more internal regions of the system than those formed in `Set 1', which of course did not include any gaseous effects. In fact, the most massive planets resulting from `Set 2' were produced around the HZ of the system, with masses between $2.2\,\mearth$ and $3.22\,\mearth$. These planets showed physical and orbital parameters similar to those associated with \gjb, which is a super-Earth with a minimum mass of $2.89 \pm 0.26\,\mearth$ orbiting around its parent star with a semi-major axis of 0.09110 $\pm\,0.00002\,\au$, and an eccentricity of 0.10$^{+0.09}_{-0.07}$ (see Table~\ref{table:sys_prop}).

Moreover, the N-body experiments of `Set 2' efficiently produced a close-in planet population within the inner edge of the HZ with masses ranging between $0.1\,\mearth$ and $1.3\,\mearth$. The physical and orbital properties of the most massive planets of this close-in population are consistent with those of \gjc, which has a minimum mass of 1.18 $\pm$ $0.16\,\mearth$, a semi-major axis of 0.036467 $\pm 0.000002\,\au$, and an orbital eccentricity of 0.17$^{+0.13}_{-0.12}$ (see Table~\ref{table:sys_prop}).

According to ‘Set 2’, which included gas dynamics, simulated planets with properties that were not dissimilar to the two known planets in the \gj\ system were formed. 
It is worth remarking that gas effects play a key role in the final planetary architecture of simulated systems. In fact, the location of the peak of the planet mass distribution at more internal regions of the system in comparison with that associated with `Set 1', and the formation of a close-in planet population within the inner edge of the HZ, are a clear consequence of incorporating type-1 migration and type-1 damping in the N-body experiments of `Set 2'. 

The right panel of Fig.~\ref{fig:figura_masas_vs_a} also allows us to derive important conclusions concerning the final water content of the resulting planets. On the one hand, all planets with masses greater than $1\,\mearth$ that survived in the HZ of the system became true water worlds. In fact, in general terms, the accretion seeds\footnote{Following \citet{Raymond2009BuildingSystem}, a planet's accretion seed is defined as the larger body in each of its collisions.} started the simulation beyond the snow line with 50$\%$ of water by mass. Throughout the $100\,\myr$ of evolution, they were impacted by planetary embryos associated with different regions of the disk, reaching final masses between $1.16\,\mearth$ and $3.22\,\mearth$, and final contents of 18\%--50\% of water by mass. On the other hand, all planets with masses less than $1\,\mearth$ that survived in the HZ had accretion seeds that started the simulation inside the snow line with 0.01\% water by mass. These planets had final masses between $0.3\,\mearth$ and $1.02\,\mearth$ and kept their very low initial fraction of water throughout their entire evolution since they only accreted planetary embryos from the inner region to the snow line. The right panel of Fig.~\ref{fig:figura_masas_vs_a} shows that \gjb\ is a potentially habitable planet, which is located in a region of the mass--semi-major axis plane associated with the presence of water worlds. According to this model, \gjb\ should have a very high water content by mass. 

\begin{figure*}
\includegraphics[width=0.97\textwidth]{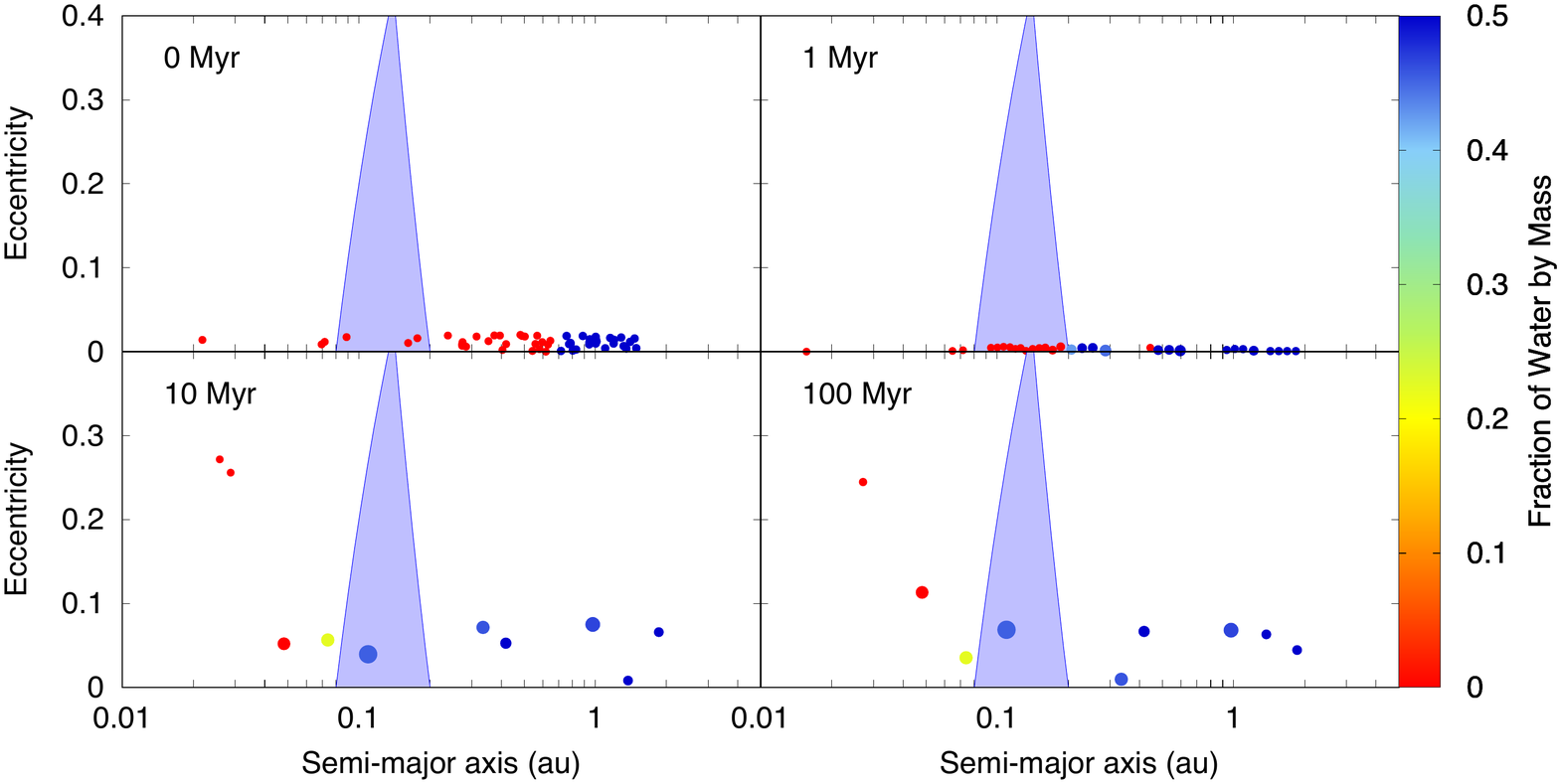}
\caption{Temporal evolution of semi-major axis as a function of the eccentricity in the 'Set 2' N-body simulations, which illustrates the most likely planetary-formation scenario.
The planetary embryos (and the resulting planets) are represented by circles, which show a colour code that indicates their fraction of water by mass. The solid blue lines illustrate curves of constant pericentric distance ($q = 0.08\,\au$) and apocentric distance ($Q = 0.2\,\au$), while the light blue shaded region represents the optimistic HZ of the system.}
\label{fig:figura_e_vs_a}
\end{figure*}



In this line of research, we suggest that planet \gjc\ should have physical properties different to those shown by \gjb\ concerning their final water contents. In fact, the right panel of Fig.~\ref{fig:figura_masas_vs_a} shows that \gjc\ is located in a region of the system where our N-body experiments produced planets with very low final water contents. 

Figure~\ref{fig:figura_e_vs_a} shows four snapshots in time of the semi-major axis--eccentricity plane of the evolution of a given N-body simulation of `Set 2', which exposes the best scenario that we found. The solid circles represent the planetary embryos (and planets) and the colour code refers to their fraction of water by mass. The solid blue lines illustrate curves of constant pericentric distance ($q = 0.08\,\au$) and apocentric distance ($Q = 0.2\,\au$), while the light blue shaded area represents the optimistic HZ of the system. Our study shows two relevant results. On the one hand, the `Set 2' simulations were able to produce planets with physical and orbital properties consistent with those associated with the planets around \gj\ \citep{astudillo2017}. In fact, this N-body experiment formed a planet analogous to GJ~273b(c) with a mass of $2.25\,(1.13)\,\mearth$, a semi-major axis of $0.1095\,(0.0482)\,\au$, and an orbital eccentricity that oscillates between $0.0007\,(0.0081)$ and $0.1\,(0.176)$ during the last $50\,\myr$ of evolution. On the other hand, the right and bottom panel of Fig.~\ref{fig:figura_e_vs_a} shows that additional planets have formed in different regions of the system after $100\,\myr$ of evolution. This result suggests that more planets with different physical and orbital properties might be orbiting around \gj.    

 It is important to remark that the final water contents of the planets formed in our simulations should be interpreted as upper limits. On the one hand, our N-body experiments treat collisions as perfect mergers, which conserve the mass and the water content of the interacting bodies. Thus, we did not account for water loss during impacts, which could be relevant for small impact angles and high velocities \citet{Dvorak2015PlanetaryPlanets}. On the other hand, our model did not include processes such as photolysis and hydrodynamic escape, which could reduce the fraction of water of the planets 
\citep{Luger2015ExtremePlanets,Johnstone2020HydrodynamicStars}. From this last consideration, GJ 273c could be certainly a dry planet since it is interior to the HZ, while GJ 273b could contain less water than that obtained in our simulations since the initial boundaries of the HZ are located in more external regions of the disk, and then, they evolve inward significantly in the first 100 Myr of evolution for a 0.3 M$_{\odot}$ M-dwarf \citep{Luger2015ExtremePlanets}. A detailed analysis about how the water loss processes mentioned in this paragraph affect the planets of our simulations is beyond the scope of this work.


\section{Observational constraints: radial velocities and photometric measurements}

\subsection{HARPS observations}
The simulations performed in Section~2 worked well in predicting the formation of two planets at the orbits of \gjb\ and \gjc. However, our results for the best model indicated that more planets might be orbiting \gj\ (see Fig.~\ref{fig:figura_e_vs_a}, panel corresponding to $100\,\myr$). The aim of this subsection is to demonstrate that, even if these planets exist, they would not be detected given the precision and cadence of the published data used to confirm \gjb\ and \gjc. The fact that \cite{Tuomi2019FrequencyNeighbourhood} is still under referee process at the time of writing, prevented us from using their data set in this experiment. Therefore, to test the hypothesis of the existence of extra planets in the system, we obtained the \harps\ RVs provided by \cite{astudillo2017} in their Table~A.1. First, we subtracted the secular acceleration following Eq.~(2) from \cite{Zechmeister2009TheUVES}. Then, we obtained the residuals to a model containing all signals present in the \harps\ data. This included, in addition to the 18.6 and $4.7\,\days$ signals of \gjb\ and \gjc, two large periodicities at $\sim$420 and $\sim700\,\days$ that the authors interpreted as stellar activity cycles. 
After having obtained the residuals, we injected RV time series representing the planets predicted by the aforementioned dynamical simulations. The goal was to check whether such signals could be detected. In particular, we defined sinusoidal models at the time epoch of \harps\ for the planet candidates represented in the last panel of Fig.~\ref{fig:figura_e_vs_a}. The periods of the sinusoids were calculated from the semi-major axes of the planets predicted by the dynamic simulations using Kepler's third law and assuming circular orbits. Additionally, we included a white noise component calculated by randomly swapping the RV data among different observing epochs and adding the resulting RVs to the sinusoidal model (via bootstrapping method). This process was repeated 1000 times for each period ($P_i$) and amplitude ($K_i$) thus obtaining 1000 simulated signals with random phases. Therefore, our goal was to progressively increase the amplitude $K_i$ and calculate the rate of simulated signals that could be detected. Such a rate is related to the minimum mass that the hypothetical planet must have in order to be detected with the cadence and precision of the \harps\ observations, assuming circular orbits. 
To this aim, we calculated the completeness ($C$) as the fraction of positive experiments ($N_{\mathrm{pos}}$) among a total of 1000 experiments ($N$) for which we recovered a signal equivalent to the sinusoid injected:
\begin{equation}
C= 100\times (N_{\mathrm{pos}}/N).
\end{equation}
To consider a recovered signal equivalent to the injected signal, i.e. a positive detection, we established that three criteria had to be satisfied: (1) the period recovered ($P_r$) had to match the injected one within the range given by the resolution frequency ($P_r = [1/(P_i+1/\mathrm{T}), 1/(P_i-1/\mathrm{T})]$; where $\mathrm{T}$ refers to the total time span of the data set), (2) the highest peak of the differential likelihood periodogram had to be above the 1$\%$-False-Alarm-Probability \footnote{For our periodograms we used a statistical estimator to the maximize the differential likelihood function ($D\ln{L} = \ln{L_{\mathrm{null}}}-\ln{L_{\mathrm{mod}}}$). In particular, it evaluated how well represented the data were when fitting the model (+1 planet, $\ln{L_{\mathrm{mod}}}$) compared to the initial hypothesis (i.e. +0 planets, $\ln{L_{\mathrm{mod}}}$). See \cite{Anglada-Escude2016ACentauri} for more details.}, and (3) the difference between the initial and recovered amplitudes had to be below the RMS of the initial residuals ($K_r-K_i<\mathrm{RMS_{\mathrm{res}}}$). 

\begin{figure}
  \includegraphics[width=\columnwidth]{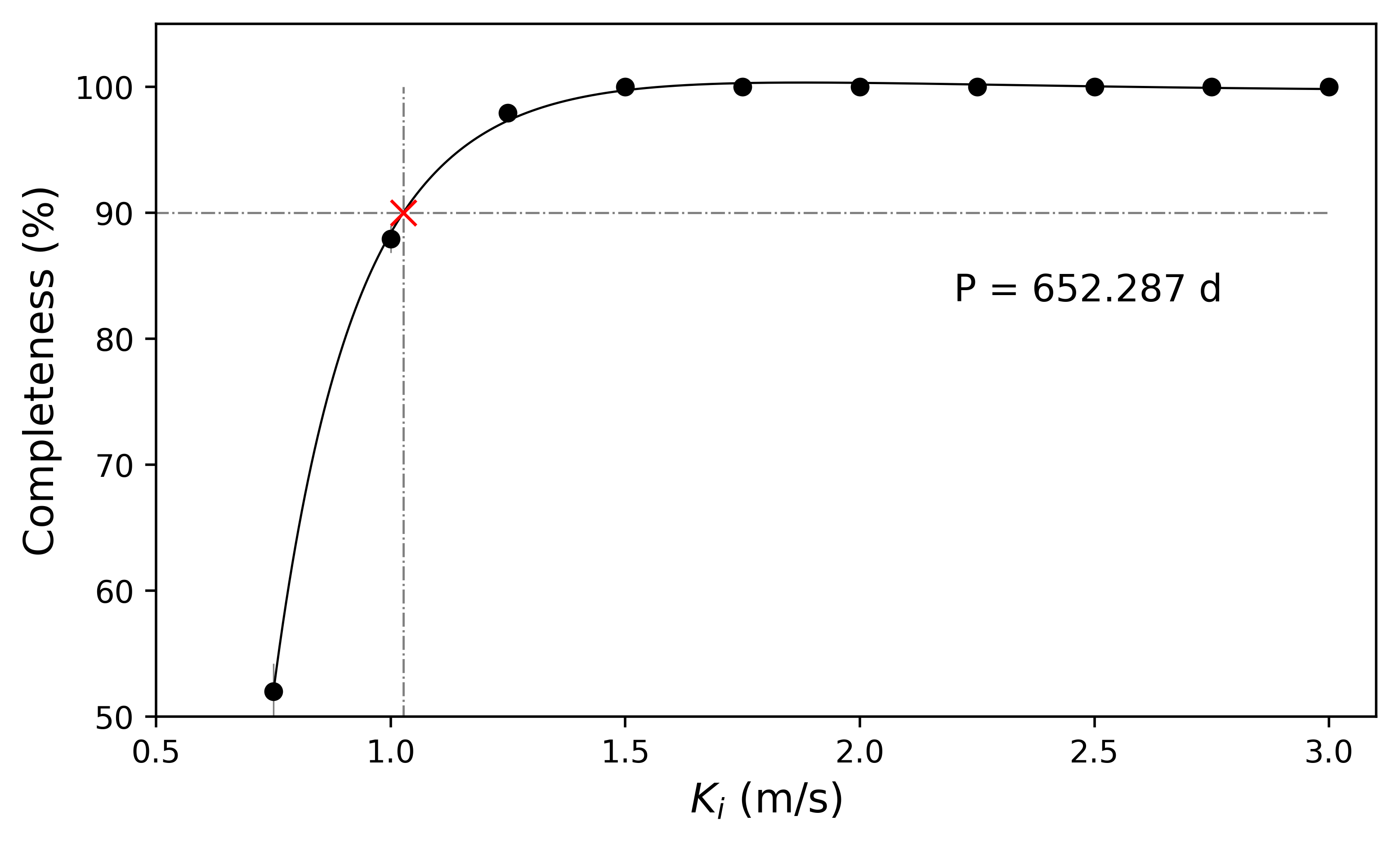}
  \caption{Percentage of well-detected sinusoids from a total of $N$=1000 experiments (i.e. completeness, $C$) versus the initial amplitude. The case of simulated sinusoids with P=652.3~d is illustrated here as an example of how the $C$ is calculate. The black line indicates the polynomial model fitted to get the initial amplitude ($K_{i}$) that corresponds to a 90\% completeness (red cross).}\label{figcompleteness}
\end{figure}

Once we obtained the completeness, we considered that signals with $C\ge90\%$ could be confidently recovered. This approach allowed us to obtain the lowest amplitude at which an RV signal could be detected, $K_{\mathrm{min}}$. Figure~\ref{figcompleteness} shows how the completeness increases with the initial amplitude for one of the cases that we tested, that is, the set of simulated sinusoids with $P=652.3\,\days$. Finally, we obtained the lower limit that the minimum mass of a planet had to reach to be detected as:
\begin{equation}
\msini = K_{\mathrm{\mathrm{min}}}\sqrt{\frac{a~M_\star}{G}},
\end{equation}
where $G$ is the gravitational constant, $M_\star$ is the mass of the star ($0.29\,\msun$), and $a$ is the semi-major axis obtained by making use of the Kepler's third law. 

\begin{figure}
  \includegraphics[width=\columnwidth]{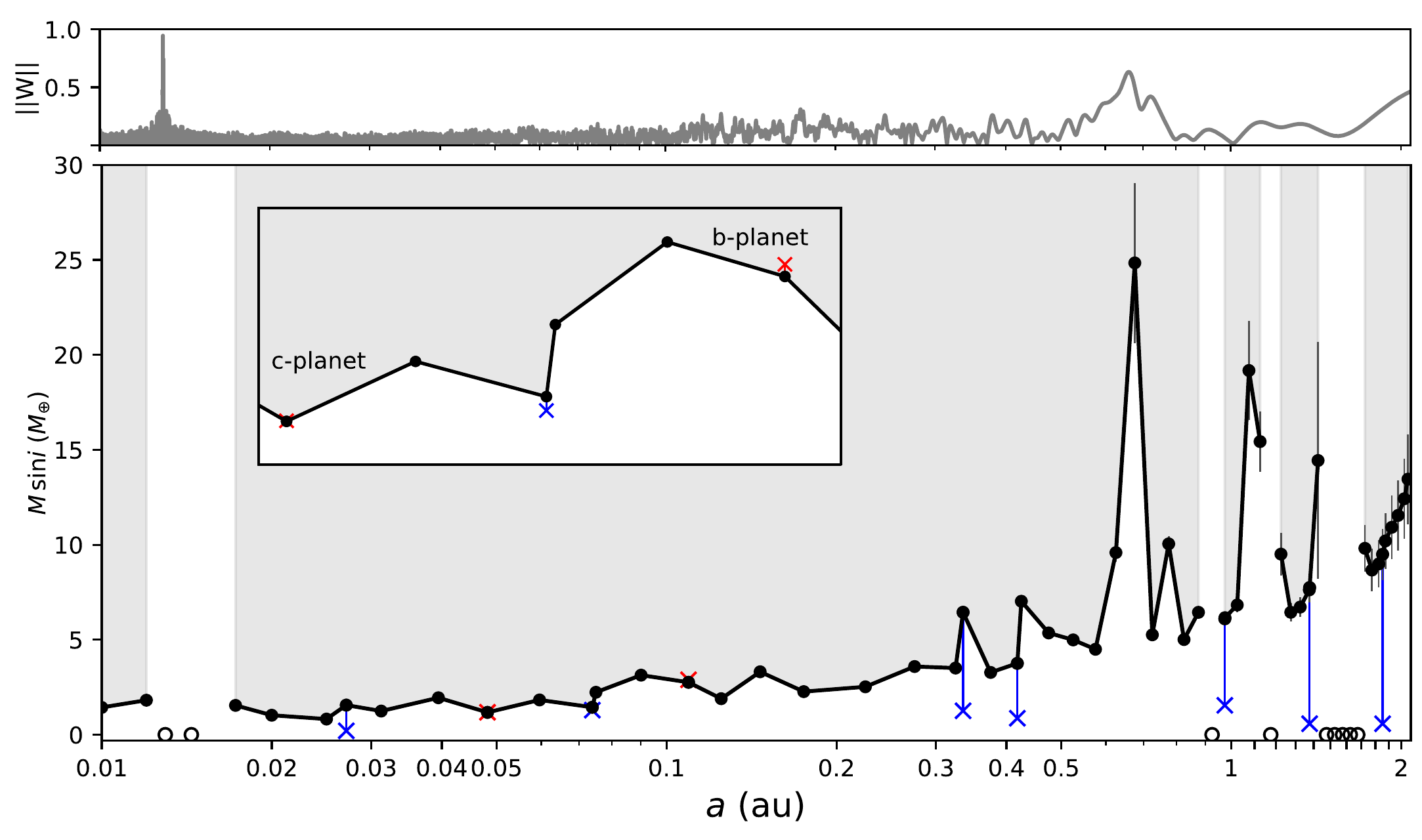}
  \caption{Minimum (lower-limit) masses of the planetary signals depending on their semi-major axes (black dots). Circular orbits were assumed. Only planets within the grey areas could be detected using the \harps\ data set that confirmed \gjb\ and \gjc. The Blue crosses, that correspond to the masses of the hypothetical planets predicted by the dynamical models (see last panel of Fig~\ref{fig:figura_e_vs_a}, are below this detection threshold. The inset shows a close-up of the \gjb\ and \gjc\ analogues predicted by the dynamical models. The minimum \gjb\ and \gjc\ masses reported by \cite{astudillo2017} are highlighted with red crosses and appear on or above the detection threshold. Open circles indicate experiments for which 90\% completeness was not reached even when running the experiments with $K_i$ up to $6.75\,\ms$. They correspond to high power regions of the spectral window shown in the top panel.}\label{figresult}
\end{figure}

The results are summarized in Fig.~\ref{figresult}. There we can see how the masses of the hypothetical planets (blue crosses) are below the detection region (grey area). On the contrary, the \gjb and \gjc analogous predicted by our dynamical simulations (red crosses) are on or above the detection limit (black line), and, as consequence, they could be detected using the RV data set. The large peaks and white gaps at $\sim0.013\,\au$ and at large semi-major axis indicate misleading regions where the temporal aliases prevent a good monitoring. This is a consequence of the spectral window shown in the upper part of the figure. 

\subsection{TESS Observations}

\tess\ (the \textbf{T}ransiting \textbf{E}xoplanet \textbf{S}urvey \textbf{S}atellite; \citealt{Ricker2014TransitingSatellite}) is currently performing a two-year survey of nearly the entire sky, with the main goal of detecting exoplanets smaller than Neptune around bright and nearby stars. It has four $24\degr\times24\degr$ field-of-view cameras, each containing four $2\,\mathrm{k}\times2\,\mathrm{k}$ CCDs, with a pixel scale of 21$^{\prime\prime}$. At the time of writing, \tess\ has discovered several planetary systems such as L 98-56 \citep{Kostov2019TheDwarf}, TOI-125 \citep{Quinn2019Near-resonanceTESS}, TOI-270 \citep{Gunther2019ATOI-270}, LP\,791-18 \citep{Crossfield2019A791-18}, and LP\,729-54 \citep{Nowak2020The729-54}  among others. 
\gj\ was observed by \tess\ in sector 7 with CCD camera 1 from 2019-Jan-07 to 2019-Feb-02 (TIC\,318686860). An alert was not issued by \tess\ for this object, due to the lack of clearly identified transits by both the QLP (Quick Look Pipeline) and the SPOC (Science Processing Operations Center). With this in mind, we performed our own search for threshold-crossing events that might not have been detected by the aforementioned automated pipelines. 

Our custom-made pipeline \sh\footnote{\sh\ code is available upon request.} (\textbf{S}earching for \textbf{H}ints of \textbf{E}xoplanets f\textbf{R}om \textbf{L}ightcurves \textbf{O}f spa\textbf{C}e-based see\textbf{K}ers) made use of the \lk\ package \citep{LightkurveCollaboration2018Lightkurve:Python} to obtain the \tess\ data of \gj\ from the \emph{NASA} Mikulski Archive for Space Telescope (MAST). We used the Pre-search Data Conditioning Simple APerture (PDC-SAP) flux given by the SPOC. First, we removed the outliers, defined as data points $>3\sigma$ above the running mean. Then, we made use of \wotan\ \citep{Hippke2019Python} to detrend the flux using the bi-weight method. This choice was based on its good speed--performance ratio compared with other methods such as Gaussian processes. The bi-weight method is a time-windowed slider, where shorter windows can efficiently remove stellar variability, but there is an associate risk of removing any actual transit signal. In order to prevent this issue, our pipeline explored twelve cases with different window sizes (see Fig.~\ref{detrend}). The minimum value was obtained by computing the transit duration (T$_{14}$) of a similar planet to the outermost one confirmed in the system, i.e., a hypothetical Earth-size planet with a period of 19~d orbiting the given star. To protect a transit of this duration, we chose a minimum window size of 3$\times T_{14}$. The maximum value explored was fixed to 20$\times T_{14}$, which seemed enough to remove the variability of \gj. For all the detrended light curves, the transit search was performed using the \tlsq\ (\tls) package, which uses an analytical transit model based on stellar parameters, and is optimized for the detection of shallow periodic transits \citep{Hippke2019TransitPlanets}. We searched for signals with periods ranging from 0.5 to 25 d, with signal-to-noise ratios (SNRs) of $\geq5$ and signal detection efficiencies (SDEs) of $\geq5$, following the vetting process described by \cite{Heller2019TransitK2}. Bearing in mind that planet \gjb\ has an orbital period of 18.6 d, since the \tess\ data span 24.45 d, including a gap of 1.67 d (TJD=1503.04-1504.71) caused by the data download during telescope apogee, this means that only a single transit may have been observed for this planet. The same situation may also apply for the outermost planets \gjd\ and \gje\, whose orbital periods exceed 400~d, which means that in the best-case scenario only a single transit might be detected for each planet in the current set of data. We did not find any signal which fulfilled the criteria of the vetting process, which agrees with the negative results yield so far by the automatic pipelines of \tess. This hints that \gj\ is not a transiting system. 

\begin{figure}
\includegraphics[width=\columnwidth]{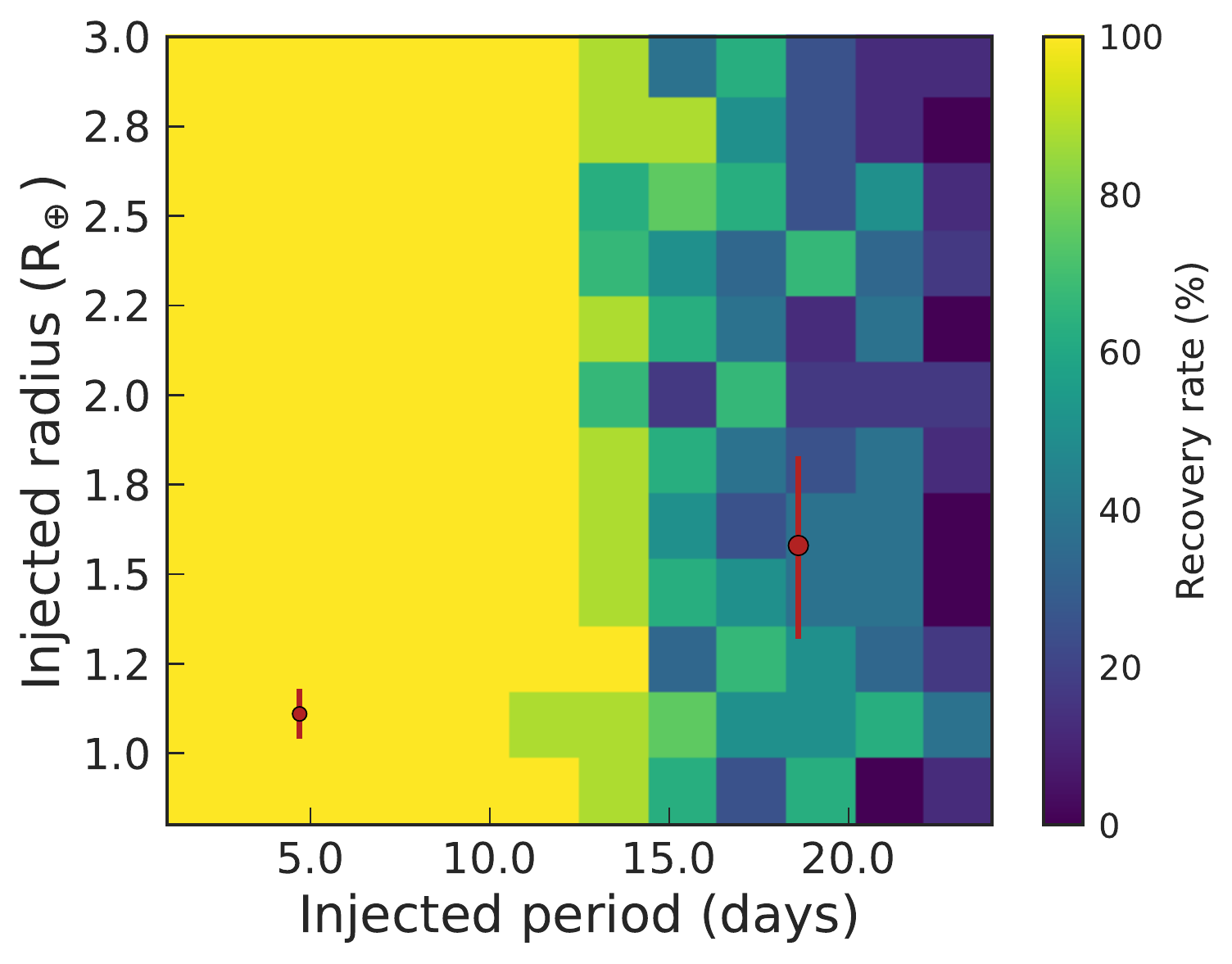}
\caption{Injection-and-recovery test performed to check the detectability of the innermost planets \gjb\ and \gjc\ in the current set of \tess\ data. 
We injected synthetic transits into the available photometric data corresponding to each planet by taking into account the uncertainties in their physical properties. We explored a total of 1080 different scenarios. We found that planet \gjc\ showed a recovery rate of 100\%, which clearly suggests that this planet is not transiting. 
On the other hand, planet \gjb\ showed a recovery rate of $\sim40\%$, implying its possible detection in future observations. We carried out two other experiments 
to check the detectability of the two outermost planets \gjd\ and \gje, finding recovery rates $<2\%$; see the main text for more details.}\label{recovery}
\end{figure}

However, in order to quantify the detectability of the planets in the existing set of data, we carried out injection-and-recovery tests. These consisted of  injecting synthetic planetary signals into the PDC-SAP flux light curve before the detrending and the \tls\ search. Since there are two confirmed planets and two candidates, with significant differences in their physical properties (i.e., the innermost planets are likely Earth- and super Earth-mass planets with orbital periods $<20\,\days$, and the outermost planets are likely mini-Neptunes with orbital periods $>400\,\days$), we performed three independent injection-and-recovery tests: (1) The first one focused on the innermost planets \gjb\ and \gjc, where we studied the $R_{\mathrm{planet}}$--$P_{\mathrm{planet}}$ parameter space in the ranges of 0.8--$3.0\,\rearth$ with steps of $0.05\,\rearth$ and 1--25 d with steps of $1\,\days$, i.e., we explored 1080 different scenarios. (2) Second, we explored the parameter space around the planet \gjd, over the radius range 1.0--4.0$\,\rearth$ with steps of $0.05\,\rearth$ and periods of 400--$425\,\days$ with steps of $1\,\days$; therefore we explored a total of 1625 scenarios. (3) For the planet \gje, we studied radii from 1.0--$4.0\,\rearth$ with steps of $0.05\,\rearth$ and periods of 530--$555\,\days$ with steps of $1\,\days$, which yielded a total of 1625 scenarios. The range of the planetary radii explored here was chosen from the dynamical arguments presented in Section~4 \footnote{The dynamic simulations in the context of the four-planet configuration yielded as 
the most likely scenario a planetary system with minimum inclinations down to 72$\degr$, i.e., composed of an Earth-mass planet, a super-Earth and two mini-Neptunes. To obtain the appropriate range of the radii, we applied the corresponding mass--radius (M--R) relationships; see Section 4 for more details.}. For simplicity, we assumed the impact parameters and eccentricities were zero. In these inject-and-recovery tests, we considered that an injected signal was recovered when a detected epoch matched the injected epoch within one hour, and if a detected period matched any half-multiple of the injected period to better than 5$\%$. As we injected the synthetic signals in the PDC-SAP light curve, the signals were not affected by the PDC-MAP systematic corrections; therefore, the detection limits should be considered as the most optimistic scenario \citep{Eisner2019PlanetOrbit}. For test (1), we found that the innermost planet \gjc\ had a recovery rate of 100\%, and the planet \gjb\ had a recovery rate of about 40\%, where these results are shown in Fig.~\ref{recovery}. On the other hand, for tests (2) and (3), the planet signals were recovered at very low rates $<2\%$. This means that a transit of planet \gjc\ can be ruled out by the \tess\ data, while transits for the other planets in the system might have been missed due to low-SNR data or because of the limited time-coverage of the data. 

\section{Dynamical evolution} 

\subsection{Stellar age}

Stellar age is a key value used to evaluate the dynamical stability of the whole planetary system. In our case, \gj\ is an M3.5V star, whose mass has been estimated to be $0.29\,\msun$ thanks to the empirical mass--luminosity relation presented by \cite{Delfosse2000AccurateRelations}. The evolution of such a low-mass star is very slow, and hence stellar evolutionary models are rather useless to establish its accurate age. According to the \parsec\ evolutionary tracks and isochrones \citep{Bressan2012PARSEC:Code,Chen2014ImprovingStars}, the stellar effective temperature T$_{\mathrm{eff}}$ and luminosity L$_{\star}$ of a $0.3\,\msun$ main sequence star vary by $\sim0.1\%$ and $\sim10\%$, respectively, over the entire age of the universe. On the other hand, it is well known that stellar age correlates with the stellar rotation rate and stellar activity, as demonstrated by several empirical relations provided in the literature. In fact, considering that \gj\ has a rotational period of $P_{\star}=99\,\days$ \citep{astudillo2017}, the gyrochronological relation presented by \cite{Barnes2010ADIAGRAMS} yields an age of $t_{\mathrm{gyro}}=8.4\,\gyr$. If, instead, we consider $\log{R'_{\mathrm{HK}}}=-5.56$ \citep{astudillo2017} as the index of chromospheric activity, the relation found by \cite{Mamajek2008ImprovedDiagnostics} yields an age t$_{\mathrm{HK}}=8.7\,\gyr$.

It is important to stress that both the empirical age-activity and gyrochronological relations are calibrated using stars (usually belonging to clusters or stellar associations) whose ages are already known thanks to isochrone fitting (i.e. evolutionary models) and whose ages are generally young. In fact, as stated by \cite{Soderblom2010TheStars}, for old stars the portion of the chromospheric signal that can be attributed to age is uncertain because it is difficult to detect and calibrate the continuing decline of activity in stars older than $\sim2\,\gyr$. In addition, there is a lack of open clusters for calibrating the gyrochronological ages of old stars. \cite{Meibom2015ACluster} confirmed a well-defined period--age relation up to $2.5\,\gyr$, which can be considered as the upper limit to which gyrochronological ages are conservatively reliable. Keeping these conservative constraints in mind, the low stellar activity level and rotational rate, and the consistency among t$_{\mathrm{gyro}}$ and t$_{\mathrm{HK}}$, all suggest an age higher than $8\,\gyr$.

\subsection{Stability of the system}

In order to explore the stability of the system, we used the Mean Exponential Growth factor of Nearby Orbits, $Y(t)$ \citep[MEGNO,][]{Cincotta1999ConditionalEntropy,Cincotta2000SimpleI,Cincotta2003PhaseOrbits}. MEGNO is a chaos index which evaluates the stability of a body's trajectory after a small perturbation and predefined initial conditions. MEGNO is closely related to the maximum Lyapunov exponent and the Lyapunov characteristic number, which are well-known powerful tools for studying the long-term evolution of Hamiltonian systems \citep{Benettin1980LyapunovApplication, Gozdziewski2001GlobalCriterion}. Therefore, MEGNO provides a clear picture of resonant structures and establishes the locations of stable and unstable orbits. To calculate MEGNO, $Y(t)$, each body's six-dimensional displacement vector, $\delta_{i}$, is considered as a dynamical variable in both position and velocity from its shadow particle, i.e., a particle with slightly perturbed initial conditions. Then, for each $\delta_{i}$ differential equation, the variational principle is applied to the trajectories of the original bodies. Then, MEGNO is straightforwardly computed from the variations as: 

\begin{equation}
 Y(t)=\frac{2}{t}\int_{t_{0}}^{t} \frac{\Arrowvert \dot{\delta}(s) \Arrowvert}{\Arrowvert \delta(s) \Arrowvert}s\,\mathrm{d}s,
\end{equation}
along with its time-average mean value: 

\begin{equation}
 \langle Y(t) \rangle=\frac{1}{t}\int_{t_{0}}^{t} Y(s)\,\mathrm{d}s
\end{equation}

\begin{figure}
\includegraphics[width=\columnwidth]{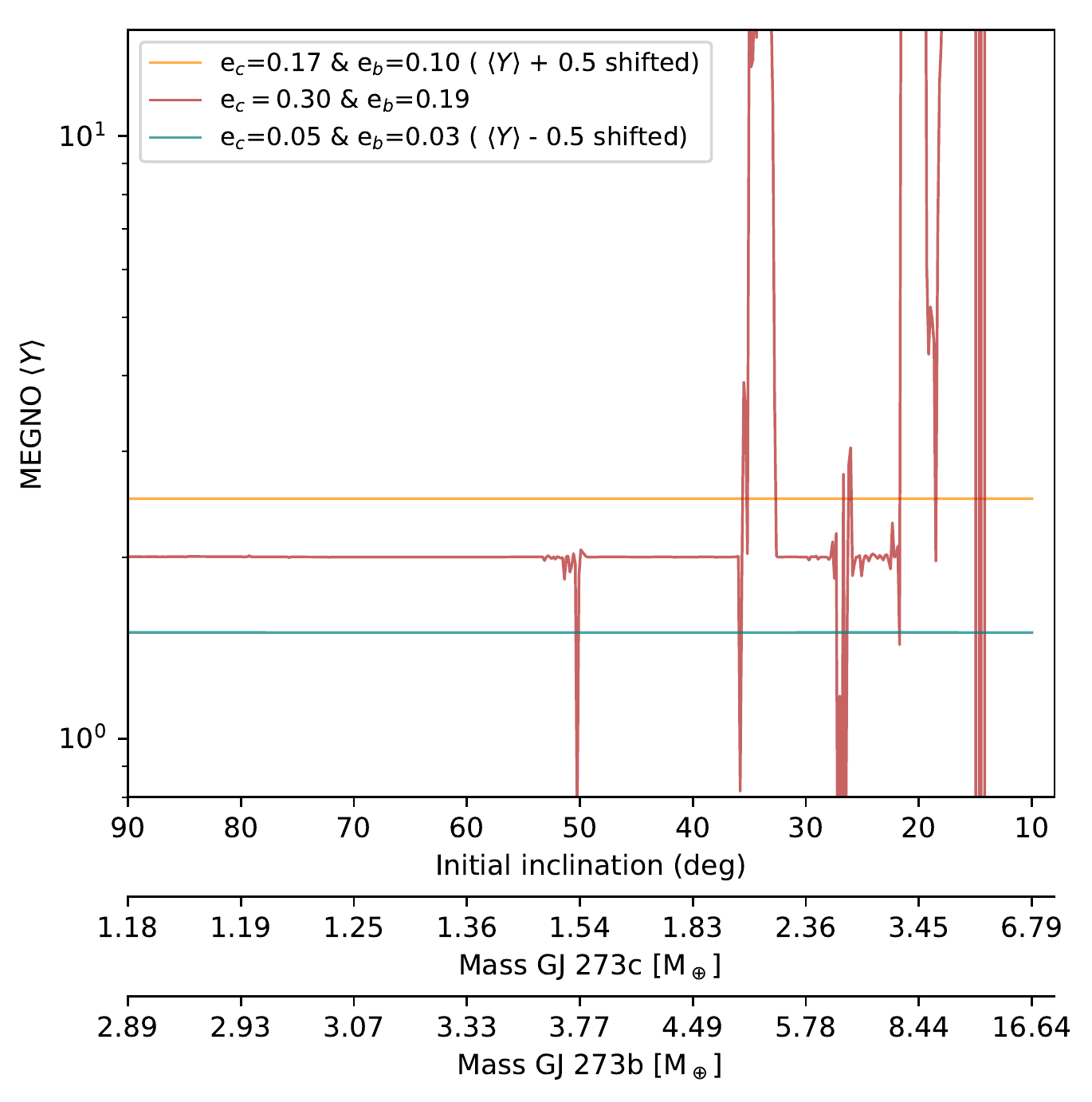}
\caption{Stability analysis based on the MEGNO parameter for different scenarios for the two-planet configuration. We explored 1000 different values of the orbital inclination ranging from 90$\degr$ to 10$\degr$, for three different sets of eccentricities corresponding to their nominal values $\pm$1$\sigma$ uncertainties:
(1) the solid-green line corresponding to $\eb=0.03$ and $\ec=0.05$; (2)
the solid-yellow line, which corresponds to $\eb=0.10$ and $\ec=0.17$; and (3) the solid-red line for $\eb=0.19$ and $\ec=0.30$. For clarity, scenarios (1) and (2) have been shifted by 0.5 units from $\langle Y(t) \rangle$=2. On the $x$-axis, for each inclination the corresponding planetary mass is shown due to the mass-inclination degeneracy.   
}\label{MegnoVsInc_e}
\end{figure}

Thanks to the time-weighted factor, the amplified stochastic behavior allows for the detection of hyperbolic regions in the time interval ($t_{0},t$). To distinguish between chaotic and quasi-periodic trajectories in phase space, we evaluated $\langle Y(t) \rangle $. Here, if $\langle Y(t) \rangle \rightarrow \infty$ for $t\rightarrow \infty$ the system is chaotic. On the other hand, if $\langle Y(t) \rangle \rightarrow 2$ for $t\rightarrow \infty$ then the motion is quasi-periodic. These criteria have been extensively used within dynamical astronomy, for both the Solar System and extrasolar planetary systems \cite[e.g.,][]{Jenkins2009FirstDesert,Hinse2015PredictingSystem,Wood2017TheRings,Horner2019TheSolution}. We used the MEGNO implementation within the N-body integrator \reb\ \citep{Rein2011REBOUND:Dynamics}, which made use of the Wisdom-Holman WHfast code \citep{Rein2015WHFast:Simulations}. This allowed us to explore the whole parameter space at a small computational cost. As a first approach we only considered the two planets already confirmed in the system, i.e., \gjb\ and \gjc. We then explored the stability of the system for different planetary eccentricities and inclinations following \cite{Jenkins2019GJ357:Simulations}. The natural trend of planetary systems is to reside in a nearly co-planar configuration, hence, we assumed planets b and c as being co-planar, and the initial longitude of the ascending node $\Omega$ for both planets was taken as zero. This choice was motivated by the results yielded by \kepler\ and the \harps\ survey, which suggest that the mutual inclinations of multi-planets systems are of the order $\lesssim$10$\degr$ \cite[see e.g.,][]{Fabrycky2014ArchitectureCandidates, Tremaine2012TheSystems,Figueira2012ComparingSystems}.
We then conducted dynamical simulations to explore the minimum values of the inclinations for planets \gjb\ and \gjc. Furthermore, their masses are dependent on their orbital inclination angles, the so called mass--inclination degeneracy of radial velocity data, which means that any uncertainties in their inclination angles will seriously affect their determined masses. Hence, exploring the minimum values of their inclination angles is equivalent to exploring the maximum values of their masses. We evaluated their inclinations 1000 times, ranging from 10$\degr$ and 90$\degr$ for three different set of eccentricities corresponding with the nominal values and $\pm1\sigma$ uncertainties given in Table~\ref{table:sys_prop}: (1) $\eb=0.03$ and $\ec=0.05$; (2) $\eb=0.10$ and $\ec=0.17$; and (3) $\eb=0.19$ and $\ec=0.30$. Our choice to limit the inclination to 10$\degr$ was motivated by geometrical arguments, which for randomly oriented systems , disfavour small values of $i$ \cite[see e.g.,][]{Dreizler2020RedGJ1061}. Furthermore, orbital inclinations close to zero (face on) disfavour the detection of planetary systems via RVs. The integration time was set to $10^{6}$ times the orbital period of the outermost planet, and the integration time-step was 5\% the orbital period of the innermost one. We found that the system was unstable when $i\lesssim50\degr$ for set (3) i.e., for the maximum eccentricity values. On the other hand, the system tolerated the full range of inclinations for sets (1) and (2) (see Fig.~\ref{MegnoVsInc_e}). This dynamical robustness impedes strong constraints on the system's inclination, and consequently on the planetary masses.             

Next, we explored the stability of the system in a four-planet configuration. In the previous analysis of the two-planet configuration we found that the system became unstable when we considered the +1$\sigma$ values of the eccentricities, i.e., set (3) for $i<50\degr$. In order to explore the effect of the inclinations on the stability of the two outermost planets, the two innermost planets were fixed to their nominal values, which yielded stable solutions. Then, we ensure that if any instability is reached is due to the new planets added to the system. With this in mind we tested three different scenarios: (1) $\ed=0.0$ and $\ee=0.0$; (2) $\ed=0.17$ and $\ee=0.03$; (3) $\ed=0.35$ and $\ee=0.23$. We found that scenarios (2) and (3) are highly unstable for the whole range of inclinations explored. Only for the case (1), for angles ranging from $90\degr$ to $\sim72\degr$, was the system stable. These provide strong constraints on the four-planet configuration, which allowed us to determine the planetary masses in the following ranges: $1.18\leq \mc\leq1.24\,\mearth$ ; $2.89\leq \mb\leq3.03\,\mearth$; $10.80\leq \md\leq11.35\,\mearth$; and $9.30\leq \me\leq9.70\,\mearth$. These results hint that for \gj\ system, if the two outermost planets are eventually confirmed, the entire system is likely composed of an Earth-mass, a super-Earth and two mini-Neptunes in the outskirts. The stability of the system is mainly controlled by the two mini-Neptunes, which should reside in nearly circular orbits. Since the planets have not been found transiting so far, their expected radii are still highly uncertain. Based on the M--R relationships for ice-rock-iron planets given by \cite{Fortney2007PlanetaryTransits}, we estimate the radius of \gjc\ to be in the range of 
$1.04\leq \rc\leq1.18\,\rearth$, considering two possible compositions: Earth-like (rock-iron as 67\%-33\%) and 100\% rocky. For \gjb\ we obtained a size range of $1.32\leq \rb\leq1.83\,\rearth$, considering compositions of either 100\% rocky, or a 50\%-50\% mix of ice-rock. These choices for the compositions of \gjc\ and \gjb\ were motivated by the results obtained in Section~2, where we found that \gjc\ is likely a very dry object, while \gjb\ was an efficient water captor at early times, and consequently it might be highly hydrated. For the outermost planets we followed the recently revisited relation given by \cite{Otegi2020RevisitedMasses}, who established an M--R diagram with two populations: small-terrestrial planets ($\rho>3.3$~g~cm$^{-3}$) and volatile-rich planets ($\rho<3.3$~g~cm$^{-3}$). This revisited M--R diagram showed that the two populations overlap in both mass (5--$25\,\mearth$) and radius (2--$3\,\rearth$), where planetary mass or radius alone cannot be used to distinguish between the two populations. This fact hints that for the case of the outermost planets in \gj\, it is not possible to distinguish among the two populations, consequently their radii are more degenerated than in the case of the inner planets. Considering both options, i.e. small-terrestrial planets and volatile-rich planets, their corresponding radius ranges are $2.00\leq \rd\leq3.68\,\rearth$ and $1.91\leq \re\leq3.35\,\rearth$, respectively.

\subsection{Tidal evolution}

 In closely-packed systems like \gj\, tidal interactions between the planets and the star may influence the evolution of the planetary parameters and their orbits. However, the time-scales differ from one parameter to other. For example, the obliquity and planetary rotational period evolve fast, while the eccentricity and semi-major axes require time-scales which may last up to the age of the host star. As a planetary orbit evolves through tidal coupling, the orbital energy is converted into tidal energy, and substantial internal heating of the planet can occur. Here, we studied the evolution of the tides in the context of the two-planet system. This choice was motivated by the fact that the outermost planets, if they exist, must not be affected by tides due their large heliocentric distances. We refer the reader to \cite{Barnes2017TidalExoplanets} and references therein for a detailed review of the existing models used to study the evolution of tides in exoplanets.
 
In order to explore the evolution of \gjb\ and \gjc\ under tidal interactions, we employed the constant time-lag model (CTL), where bodies were modelled as a weakly viscous fluid 
\cite[see e.g.,][]{Mignard1979TheI,Hut1981TidalSystems.,Eggleton1998TheFriction,Leconte2010IsEccentricity}. We made use of the implementation of this model in \merct\ \citep{Bolmont2015Mercury-T:Kepler-62} and \possi\
\citep{Blanco-Cuaresma2017StudyingRust}. Our planetary formation model favored small planets in the range of their uncertainties, i.e., \gjc\ in the range of 1.18--$2.50\,\mearth$ and \gjb\ in 2.89--$6.00\,\mearth$. Hence, \gjc\ may be considered an Earth-mass/rocky planet, while \gjc\ might be a super-Earth or a mini-Neptune composed of a mix of rock and ice. With these in mind, for \gjc\ we assumed the product of the potential Love number of degree 2 and a time-lag corresponding to Earth's value of $k_{2,\oplus}\Delta\tau_{\oplus}=213\,\mathrm{s}$ \citep{NerondeSurgy1997OnEarth}. Furthermore, we assumed that the potential Love number of degree 2 and the fluid number were equal. For rocky-icy planets, the dissipation factor strongly depends on the ratio of rock to ice, where a larger amount of ice results in a larger dissipation factor. Unfortunately, these values are unknown. For example for planets with 100\% ice the dissipation factor is considered to be larger than Earth's by up to 10 times. For 100\% rocky planets, this factor is assumed to be roughly 0.1--0.5 Earth's value. Since from our formation models the planet \gjb\ seems to be an efficient water captor at early times, we considered that the ratio of rock and ice was skewed towards ice-rich. Hence, for \gjb\ we assumed 3$\times k_{2,\oplus}\Delta\tau_{\oplus}$ \citep{Bolmont2015Mercury-T:Kepler-62,Bolmont2014FormationSystem,McCarthy2013PlanetaryProperties}. This choice, despite being made without taking into account the results found in the four-planet configuration, it is still in agreement with 
the results of the latter configuration, where it was found that \gjc\ is likely an Earth-mass planet and \gjb\ is a super-Earth. A detailed study of the planetary compositions would be desirable in order to better understand the real nature of the planets and the validity of the values assumed here.

\subsubsection{Obliquity and planetary rotational periods}

\begin{figure}
\includegraphics[width=\columnwidth]{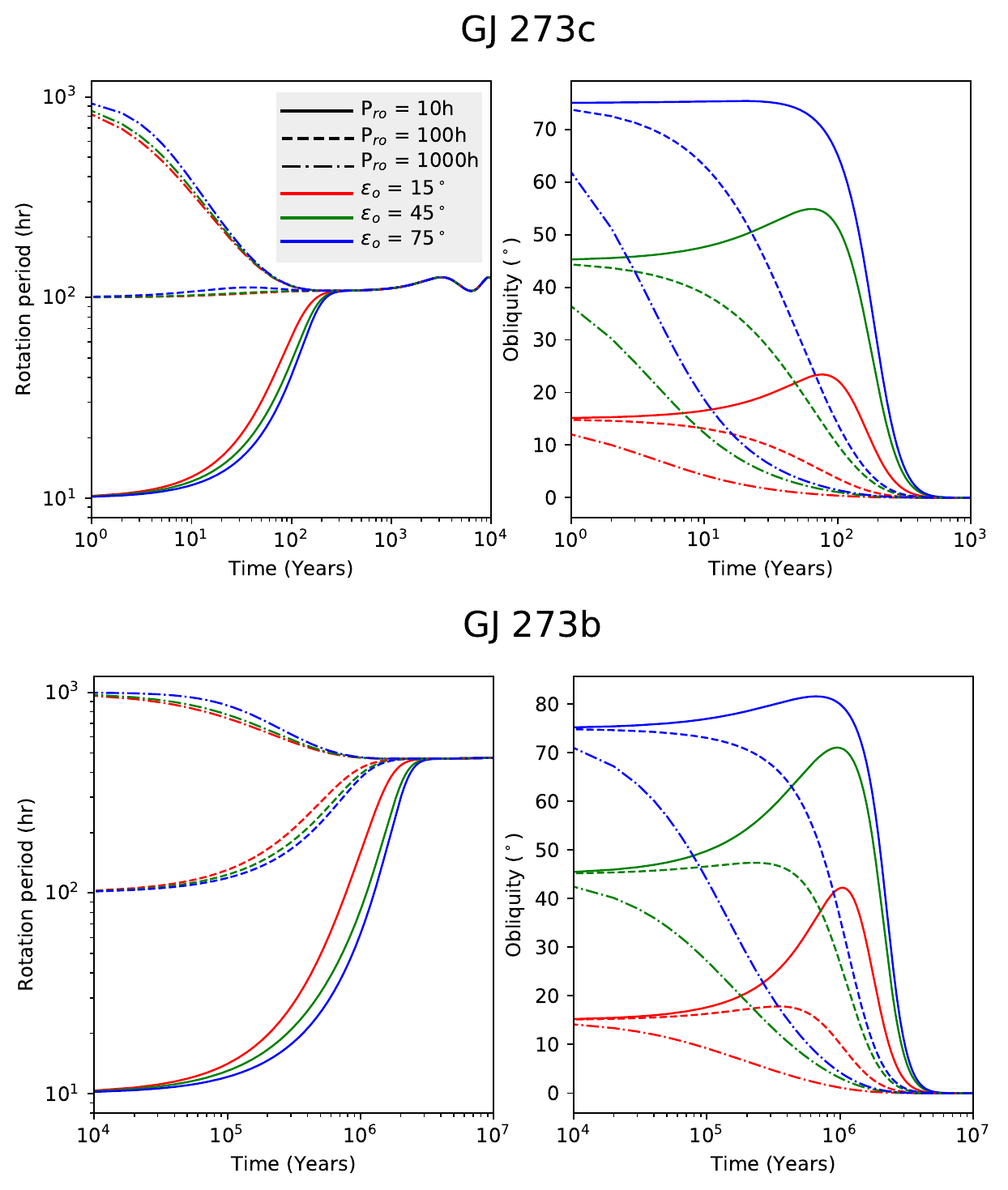}
\caption{Obliquity and rotational period evolution due to the effect of tides for the innermost planets \gjb\ and \gjc. Each set of line-style curves represents a different initial spin-rate, and each set of coloured curves represents a different initial obliquity. It is shown that for any possible combination of initial conditions, in a short-time scale compared with the age of the star both planets reached a pseudo-rotational state.}\label{rot_obli}
\end{figure}

In order to explore the evolution of the obliquity ($\epsilon$) and the rotational periods of the planets ($\mathrm{P}_{\mathrm{rot}}$), we performed a set of simulations with different initial conditions: planetary rotational periods of $10\,\hours$, $100\,\hours$, and $1000\,\hours$, combined with obliquities of $15\degr$, $45\degr$ and $75\degr$. The rest of the values were fixed to the nominal values in Table~\ref{table:sys_prop}. We found that planet \gjc\ evolved over a short time-scale ranging between $10^{4}$ to $10^{5}\,\yr$ to a  state with  $\epsilon \rightarrow 0\degr$ and $P_{\mathrm{rot}}=113\,\hours$. For planet \gjb, the evolution was slower and it needed a time-scale of about $10^{6}-10^{7}\,\yr$ to evolve towards a state with a $\epsilon \rightarrow 0\degr$ and $P_{\mathrm{rot}}=447\,\hours$. The results of this set of simulations are displayed in Fig.~\ref{rot_obli}. Since \gj\ is much older than these time-scales, we conclude that the planets of the system are in pseudo-rotational state, i.e., the planet's rotational spin is oriented with the star's (the obliquity is zero) and it reached a constant rotational period. That is the planets are tidally-locked and fairly well aligned with the host star. However, it is also important to note that the presence of a relatively large moon might perturb the pseudo-rotational state, thereby locking the obliquity into another value or generating chaotic fluctuations \citep[see e.g., ][]{Laskar1993ThePlanets,Lissauer2012ObliquityEarth}.   

\subsubsection{Eccentricity, semi-major axes and tidal heating}

\begin{figure}
\includegraphics[width=\columnwidth]{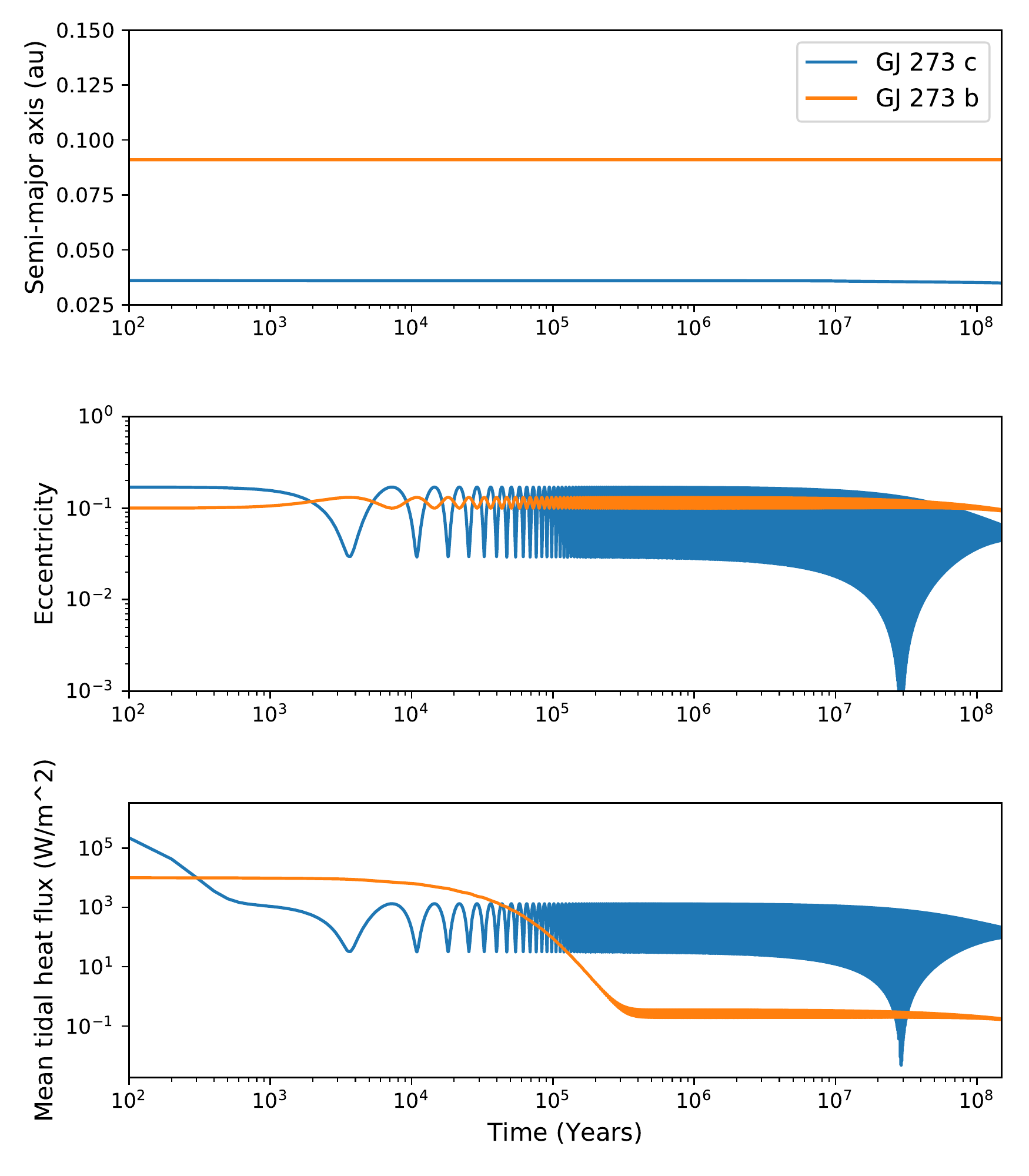}
\caption{Temporal evolution of the semi-major axis (top panel), eccentricity (mid-panel) and mean tidal heating (bottom-panel) for the innermost planets \gjb\ and \gjc\ in the system due to the effects of tides.}\label{orbit}
\end{figure}

To study the evolution of the orbits we ran another suite of simulations, with integrations up to $\sim10^8\,\yr$, assuming the nominal values given in Table~\ref{table:sys_prop}, and both planets had starting obliquities of $15\degr$, and rotational periods of 10~hr. From the previous analysis, we note that these choices do not make a substantial differences since the time-scales for the evolution of these parameters are much shorter. The results of these simulations are displayed in Fig.~\ref{orbit}. The semi-major axes 
of the planets do not show any remarkable evolution, which agrees with results presented by other authors where it has been shown that the evolution of the semi-major axis due to the effect of tides might last as long as the life of the star \cite[see e.g.,][]{Bolmont2015Mercury-T:Kepler-62, Bolmont2014FormationSystem}. 

Regarding the eccentricities, for the inner most planet, \gjc\ shrinks from 0.17 to 0.12 at the end of the integration. For planet \gjb\ the eccentricity keeps its initial value of 0.10. This hints that the tides are not strong enough to provoke the circularization of the orbits in the integration time lapse considered in this study. From this slow evolution we can infer that
the circularization spans from hundreds to thousands millions of years. 

Concerning the tidal heating, once a pseudo-rotational state is reached by a planet, where from the previous sub-section we found happens over a short time scale of $<10^{7}\,\yr$, tidal heating remains at play while the orbits are eccentric. We have shown that the period of circularization lasts, at least, several $10^{8}\,\yr$. A similar result was obtained by \cite{Barnes2017TidalExoplanets} for Proxima Centauri-b, who also used the CTL model. The planet was found to reach the tidal-locking configuration in a time-scale of $10^{4}$--$10^{5}\,\yr$, but the orbit was likely non-circularized. 
Taken together, the results found here suggest that: (1) tidal heating acts over long-time scales of at least several $\sim10^{8}\,\yr$ or more, and (2) it is likely that the planets are tidally locked, but the same side may not always face the star. For the innermost planet, the tides add significant extra heating which oscillates in the range of 100-$1000\,\mathrm{Wm}^{-2}$. On the other hand, \gjb\ received a tidal heating of $\sim0.1\,\mathrm{Wm}^{-2}$.

\section{Minor-body reservoir analogues}

 In the Solar System, minor bodies are the building blocks of planets. It is extensively accepted that they are the leftovers of planetary formation, which, through impacts, played an important role in the hydration process and the delivery of a variety of organic volatiles to the early Earth \cite[see e.g.,][]{Owen1995CometsAtmospheres,Morbidelli2000SourceEarth,Horner2009DifferencesRatios,Hartogh2011Ocean-like2,Snodgrass2017TheSystem}. This hints that minor bodies might have played a key role in the emergence of life on Earth \cite[see e.g,][]{Oberbeck1989ImpactsLife,Oberbeck1989EstimatesLife,Nisbet2001TheLife}. On the other hand, it is also important to consider that a too-high impact flux of minor bodies might erode planetary atmospheres, thereby provoking oceanic evaporation, sterilizing planetary surfaces, and even causing mass extinctions \cite[see e.g.,][]{Alvarez1980ExtraterrestrialExtinction,Chyba1993TheUncertainties,Becker2001ImpactFullerenes,Ryder2002MassEarth,Hull2020OnBoundary}. Regarding exoplanetary systems, two populations of comets orbiting $\beta$-Pictoris have been identified \citep{Kiefer2014TwoSystem}. In addition, there is strong evidence of falling evaporating bodies, i.e. exocomets, in other young systems such as Fomalhaut and HD\,172555 \cite[see e.g.,][]{Kiefer2014Exocomets172555,Matra2017DetectionComets}. There are also high-contrast, spatially resolved observations which showed an excess in the near infrared that was attributed to the presence of exozodiacal light in the main-sequence stars, whose origin may have been the same as that producing the zodiacal cloud in the Solar System: minor bodies' dust particles coming from asteroid and cometary activity and/or disruptions \citep[see e.g.,][]{Nesvorny2010CometaryDisks,Absil2013ACHARA/FLUOR,Sezestre2019HotOrigin}. Furthermore, the recent discoveries of the interstellar objects 1I/2017~U1 \cite[see e.g.,][]{Meech2017AAsteroid, Jewitt2017InterstellarTelescopes} and 2I/Borisov \cite[see e.g.,][]{Fitzsimmons2019Detection2I/Borisov,deLeon2019InterstellarGTC}, clearly hint at the existence of minor bodies in other planetary systems, which can be ejected presumably by some kind of dynamical perturbation, similar to what happens in the Solar System. Moreover, \cite{Ballesteros2018KIC2021} proposed that a swarm of Trojan-like asteroids could explain the mysterious behaviour of KIC\,8462852 \cite[see e.g.,][]{Boyajian2016PlanetFlux,Boyajian2018The8462852}. 
 Despite being a very new and emerging field\footnote{From 13 May through May 17, 2019 at Leiden in the Netherlands, an exocomet conference was held for first time: a Lorentz workshop entitle \emph{ExoComets--Understanding the composition of Planetary Building Blocks} which brought together the Solar System comet community, exoplanet researchers, and astrobiologists.}, the formation of minor bodies together with planets around host stars seems an evidence, and therefore, almost every planetary system may harbour such remnants, which may reside in stable orbits after the planetary-formation process. In this context we explored the potential existence and location of regions in \gj\ where minor bodies could reside in stable orbits for long periods of time. In our Solar System, these stable regions are the minor-body reservoirs, such as the main asteroid belt (MAB) \cite[see e.g.,][]{DeMeo2015TheBelt, Morbidelli2015TheBelt}, the Kuiper belt (KB) \cite[see e.g.,][]{Jewitt1993DiscoveryQB1,Brown2001TheBelt,Morbidelli2020KuiperEvolution} and the Oort cloud (OC) \cite[see e.g.,][]{Dones2004OortDynamics,Fernandez2005CometsRelevance}. For a full review of the characteristics and properties of these structures we refer the reader to \cite{deLeon2018TheSystem} and references therein. In this current analysis, we investigated the similarities between \gj\ and our Solar System.  

To perform this study we made use of the MEGNO criterion, previously described in Section~2. We carried out a stability analysis by introducing a mass-less particle, such as a comet or an asteroid, into the \gj\ system. Initially we only considered the two-planet configuration. The particle was then evaluated in the $a$--$e$ parameter space using initial values of $a_{0}\in[0.0, 0.2]\,\au$ and $e_{0}\in[0.0, 0.99]$. For simplicity, we considered the particle as coplanar, i.e. $i=0\degr$, meaning we only investigated moderately flat structures such as MAB and KB analogues. This choice was motivated by the complex formation process of the OC, which implies the migrations of giant planets in early ages of the Solar System. Due to the many uncertainties in the planetary parameters for the two-planet configuration, we repeated this analysis for different scenarios by considering the maximum and minimum values of their mass and eccentricity, $e$, as constrained by the dynamic considerations presented in Section~4, but we kept their semi-major axis, $a$, and mean longitude, $\lambda$, constant. 

\begin{figure*}
  \centering 
  \includegraphics[width=0.75\textwidth]{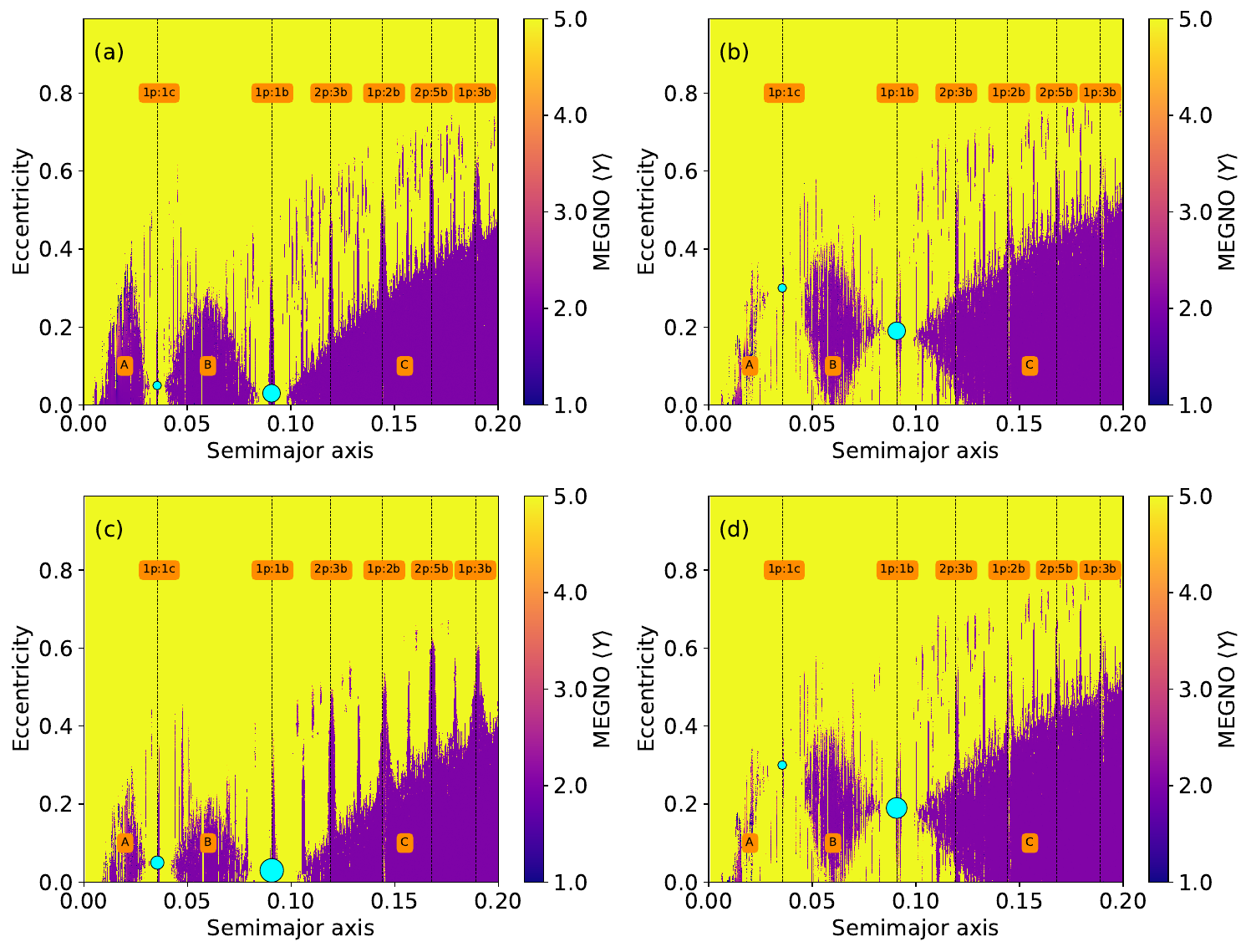}
  \caption{MEGNO maps in the $a$--$e$ parameter space for the two-planet configuration for four different scenarios (as shown in Table~\ref{table:mb}). The dark-purple areas correspond to stable orbits where minor bodies may reside. On the other hand, yellow areas correspond to unstable regions from which a minor body would be expelled. Three main stable regions are found: A, B, and C. Vertical dashed-lines represent the MMRs found in the system between the minor bodies and the planets \gjb\ and \gjc.}\label{mba}
\end{figure*}

We tested four extreme configurations (see Table~\ref{table:mb}), which covered all the possible parameter variations. The integration time was set to $10^{4}\times$ the orbital period of the outermost body in the simulation. The time-step for the integration was set to 5\% of the orbital period of the innermost body. The size of the parameter space explored was $500\times500$ pixels, i.e., we studied the $a$--$e$ space up to 250\,000 times. In order to save computational time, we prematurely ended the simulation when $\langle Y(t) \rangle \textgreater 5$, or when a collision of two bodies or the ejection of a particle took place. For all these chaotic scenarios we assigned a value of $\langle Y(t) \rangle = 5$. We found that there are three main regions (Fig.~\ref{mba}) where a particle remained stable. These are labelled as Regions A, B and C. Region A is placed between the host star and the inner planet \gjb, from $\sim0.01\,\au$ to $\sim0.03\,\au$ and $e\lesssim 0.4$. Region B is situated between the two planets, from $\sim0.040\,\au$ to $\sim0.07\,\au$, and $e< 0.3$. The last stable region, C, is the largest one, which is located beyond the outermost planet from $\sim0.1\,\au$ to $\sim0.2\,\au$, where $e$ increases progressively from 0 up to 0.4. In the Solar System, the MAB is a very stable region located between Mars and Jupiter, which contains roughly $3\times\,10^{24}\,\mathrm{g}$ ($\sim4\%$ mass of the Moon) of material spread among countless bodies with a size distribution given by a power law with an incremental size index between -2.5 and -3, most of whom have eccentricities less than 0.4 and inclinations below $20\degr$ \citep[see e. g., ][]{deLeon2018TheSystem}. The objects in this region are composed mainly of silicates, metal and carbon. Since Region B is inner to the snow line of the system ($\sim0.65\,\au$), we might consider this as a hot-MAB analogue, where minor bodies may reside, trapped in between the two planets. This region, while being sculpted slightly differently in each scenario, robustly remained in the system throughout all the possible configurations explored, which hints that it is a very suitable place where minor bodies may exist. Furthermore, our planetary formation models presented in Section~2 revealed that bodies in this region are very dry, similar to what is found in our MAB. On the other hand, Region A suffered dramatic changes depending on the configuration considered, even disappearing when the maximum eccentricities were considered, i.e. scenarios (b) and (d). Therefore, the existence of objects in region A seems less possible, but perhaps mildly plausible. However, in a more realistic scenario, the tidal interactions with the host star might prevent the accumulation of minor bodies in this region. In the Solar System there is no population of objects inhabiting stable orbits between the innermost planet, Mercury, and the Sun, which hints that this structure does not have any known analogue. Finally, we identified a region which expanded from the outermost planet to the edge of the system considered at $0.2\,\au$, i.e. Region C. This region changed slightly in each scenario, but it remained in all of them. Its shape was sculpted by planet \gjb, which was also responsible for several spines of stability, which corresponded to mean motion mesonances (MMRs). The most evident MMRs have been labelled in Fig.~\ref{mba}, but many others with lower intensities were also present. Similar structures were also found in Kepler-47 by \cite{Hinse2015PredictingSystem}. In order to identify these MMRs, we made use of the \atlas code \footnote{\url{http://www.fisica.edu.uy/~gallardo/atlas/2bmmr.html}}, which allowed us to place resonances between a planet of mass $m_{\mathrm{p}}$ and a mass-less particle \citep{Gallardo2006AtlasSystem}. Region C seems to be similar to the reservoir present in the outskirts of the Solar System, i.e., the KB. In the Solar System, this region is differentiated into two areas: the trans-Neptunian belt and the scattered disk (SD). The first contains objects with eccentricities up to 0.2, while in the SD the bodies may have eccentricities up to 0.5; in both cases the inclinations are mostly below 40$\degr$ \citep[see e. g., ][]{deLeon2018TheSystem}. The objects contained in this region are largely composed of ices, such as water, nitrogen, or methane, mixed with silicates and refractory non-volatiles species. From our planetary formation models in Section~2, we found that Region C is composed mainly by high-hydrated objects, what could be considered also as volatile-rich. Since this region is also inner to the snow line, it might be a hot-KB analogue. Among all the resonances found in Region C, it is interesting to remark the 2:3 MMR with \gjb, which is analogous to the well-known 2:3 resonance of minor bodies with Neptune, which is one of the most populated resonances in the Solar System, where the dwarf planet Pluto and the Plutinos family orbit. It has been suggested that trans-Neptunian objects larger than 500~km may have a rocky core, an icy mantle, and a volatile-rich crust \citep{McKinnon2008StructurePlanets}. In fact, the NASA's New Horizon spacecraft visited the Pluto--Charon system, it found strong indications of the presence of a subsurface ocean on Pluto; such oceans may also be present in others dwarf planets \citep{Moore2016TheHorizons}. In this context, it is interesting to note that for \gj, this hot-KB analogue resides in the HZ of the system.

\begin{figure}
  \includegraphics[width=\columnwidth]{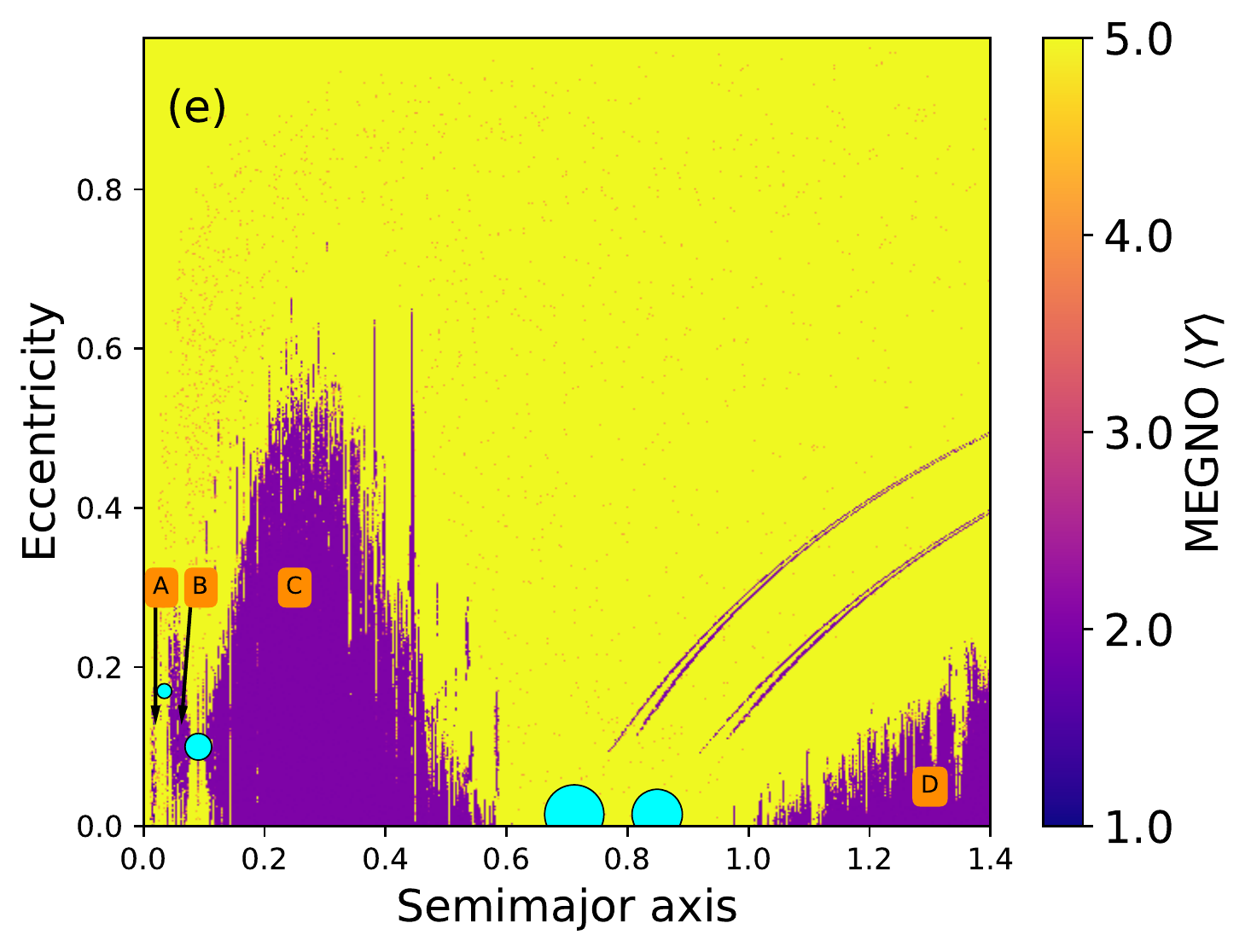}
  \caption{MEGNO map in the $a$--$e$ parameter space for the four-planet configuration, as shown in Table~\ref{table:mb}. The colour scheme is the same as given in Fig~\ref{mba}. }\label{4pla}
\end{figure}

In the second step of our analysis, we considered the four-planet configuration. In this case, as shown in Section~4, we found that the system remained stable only for values of inclinations ranging from 90$\degr$ to $\sim$72$\degr$; hence the masses of the planets were highly constrained. Also we found that the two outermost planets likely resided in near-circular orbits. Therefore, since the planetary parameters were well constrained, for the four-planet configuration we only studied one scenario (Table~\ref{table:mb}). We followed the same strategy described for the two-planet configuration but extended the initial semi-major axis range to $1.4\,\au$. We assumed an intermediate value of the inclination of $80\degr$. A stability map of this configuration is displayed in Fig.~\ref{4pla}. We found that the innermost Region A was almost removed in this scenario, similar to that found for scenarios (b) and (d) shown in Fig.~\ref{mba}. Region B remained well established, which lends credence to the conclusion that is a very suitable region where minor bodies may reside. On the other hand, Region C, which we considered to be a hot-KB analogue in the two-planet configuration, consisted of a large stable region spanning from 0.1 to $0.6\,\au$, with a maximum eccentricity of $\sim0.55$ at $0.3\,\au$. This structure may be considered as the second hot-MBA analogue in the system. From $\sim$0.55 to $\sim1.0\,\au$ there was an empty region where minor bodies would be expelled easily, and from $\sim1.0\,\au$ to the edge of the simulation 
there was a stable region which may be a KB analogue, which we have labelled as Region D. Therefore, in the four-planet configuration, \gjb\ resided in between the two hot-MBA analogues. Regions C and D were sculpted by the pair of mini-Neptunes in the outskirts, which may have likely provoked long-terms perturbations which altered the state of the minor bodies hosted therein.

This technique may be used not only to identify the location(s) of minor-body reservoirs such as MAB or KB analogues in exoplanetary systems, but also to predict the locations and ascertain the characteristics of debris disks. An example is the analysis of the dynamic environment of HR\,8799 made by \cite{Contro2016Modelling8799}, where the authors modelled the innermost debris disk and concluded that it exhibited similar features as our MAB, including the presence of Kirkwood gaps. Indeed, in the case of \gj, the system might harbour debris disks in these stable locations. However, the \gj\ system is old and it is expected that the primordial dust and gas have been removed a long time ago, as happened in our own Solar System. If in the future infrared or millimetre observations of this system reveal the presence of debris disks, one may consider that these regions are collisionally active, i.e., collisions between minor bodies happen very often and produce vast quantities of dust large and common enough to prevent its dissipation via a variety of mechanisms such as Poynting-Robertson drag or radiation pressure, which act on short time-scales \citep[see e.g.,][]{Wyatt1950TheOrbits,Wyatt2011DebrisProcesses}.

\begin{table}
\caption{Scenarios considered in Section~5}             
\label{table:mb}      
\centering 
\begin{tabular}{cccccc}     
\hline\hline       
\noalign{\smallskip}                     
Sce. & Planet & $a$~(au) & $e$ & $i$~($\degr$) & $M\,(\mearth)$ \\
\hline
\noalign{\smallskip}
\multirow{2}{*}{(a)}  & b & 0.091 & 0.03 & \multirow{2}{*}{90} & 2.89 \\
                      & c & 0.036 & 0.05 &  & 1.18 \\
\noalign{\smallskip}
\hline                                                                 
\noalign{\smallskip}
\multirow{2}{*}{(b)} & b & 0.091 & 0.19 & \multirow{2}{*}{90} & 2.89 \\
                     & c & 0.036 & 0.30 &  & 1.18 \\
 
\noalign{\smallskip}
\hline                                                                 
\noalign{\smallskip}
\multirow{2}{*}{(c)} & b & 0.091 & 0.03 & \multirow{2}{*}{10} & 16.64 \\
                     & c & 0.036 & 0.05 &   & 6.79 \\
\noalign{\smallskip}
\hline                                                                 
\noalign{\smallskip}
\multirow{2}{*}{(d)} & b & 0.091 & 0.19 & \multirow{2}{*}{50} & 3.77 \\
                     & c & 0.036 & 0.30 &  & 1.54 \\ 
            
\noalign{\smallskip}
\hline                                                                 
\noalign{\smallskip}
\multirow{4}{*}{(e)} & b & 0.091 & 0.10 & \multirow{4}{*}{80} & 2.93 \\
                     & c & 0.036 & 0.17 &  & 1.19 \\ 
                     & d & 0.712 & 0.0 &  & 10.96 \\
                     & e & 0.849 & 0.0 &  & 9.44 \\
                     
\noalign{\smallskip} 
\hline  

\end{tabular}
 \tablefoot{Set of scenarios used to study the minor-body reservoirs for both two- and four-planet configuration. The result of the simulation of each scenario is displayed in Fig.~\ref{mba} and Fig.~\ref{4pla}.}\\  
            
\end{table}

\section{Habitability of \gj}

Long has the importance of M dwarfs in the search for life-bearing planets been debated. It was suggested that, for many aspects, planets orbiting in the liquid-water HZs of these stars might provide stable environment for life, especially just after their first $0.5-1.0\,\gyr$, which consists of a period of intense radiation which could erode and damage planetary atmospheres. This, along with the feasibility of transit detection compared with more massive stars, has prioritized M-dwarf's planets as targets in the search of other biospheres beyond the Solar System \cite[see e.g.,][]{Scalo2007MDetection,Yang2013STABILIZINGPLANETS,Shields2016TheStars,OMalley-James2019LessonsSystems}.    

In this context, the \gj\ system is particularly interesting because it is one of the closest planetary systems to Earth hosted by an M dwarf, and which contains a confirmed exoplanet in the inner region of the HZ. Indeed, the first factor to consider whether a planet is potentially habitable is its distance to the host star, in which a terrestrial-mass planet with an Earth-like atmospheric composition (CO$_{2}$, H$_{2}$O and N$_{2}$) can sustain liquid water on its surface. Following \cite{Kopparapu2013HabitableEstimates,Kopparapu2014HABITABLEMASS}, the optimistic HZ in the \gj\ system spans from 0.08 to $0.2\,\au$, which establishes that \gjb\ is close to the inner border, but still inside it.

 However, many different factors play important roles when describing the potential habitability of an exoplanet, and for a detailed discussion we refer the reader to \cite{Meadows2018FactorsHabitability} and references therein. Unfortunately, it is not possible to address each of these factors at the same time. 
 
In this particular study, from the planetary formation modelling performed in Section~2, we found that \gjb\ could have been an efficient water captor during its first $100\,\myr$. Since we are considering perfect merger collisions, the water mass fraction found for this planet is an upper limit, and it is likely that the real content of water is less, but still significant, which implies a certain level of hydration.  
 
From the dynamic analysis performed in Section~4, we found that \gjb, if the four-planet configuration is confirmed, would be compatible with a super-Earth with a mass range of $2.89\leq \mb\leq3.03\,\mearth$, a size of $1.32\leq \rb\leq1.83\,\rearth$, and it resides in a stable orbit. The orbital parameters of a given exoplanet, such as its semi-major axis, eccentricity and obliquity, govern the radiation received by the planet along its orbit, which impacts its planetary climate \cite[see e.g.,][]{Spiegel2009HABITABLEOBLIQUITY,Armstrong2014EffectsExoplanets,Shields2016TheKepler-62f}. In the work by \cite{Kane2017ObliquityExoplanets}, the authors described the maximum flux received by a planet as a function of the latitude $\beta$, the orbital eccentricity, the orbital phase $\phi$, and the obliquity $\epsilon$, as follows: 
\begin{equation}\label{eq:F}
    F=\frac{L_{\star}}{4\pi r^{2}}\cos|~\beta-\delta~|,
\end{equation}
where L$_{\star}$ is the stellar luminosity, and $\delta$ is the stellar declination given by: 

\begin{equation}\label{eq:delta}
    \delta=\epsilon \cos[2\pi(\phi-\Delta \phi)], 
\end{equation}
for which $\Delta \phi$ is the offset in phase between periastron and the highest solar declination in the Northern hemisphere. As we showed in Section~4, due of the action of tides, \gjb\ is likely in a pseudo-rotational state, i.e., the obliquity is zero and it reached a constant rotational period. However, the planet's orbit might not be completely circularized yet. Consequently, Eq.~\ref{eq:F} and ~\ref{eq:delta} can be simplified as: 

\begin{equation}\label{eq:F_s}
    \delta=0 \rightarrow F=\frac{L_{\star}}{4\pi r^{2}}\cos|~\beta~|
\end{equation}

This means that the orbital phase is no longer relevant, and the distribution of flux along the latitude is the same for the whole orbit, which depends only on the planet's eccentricity. Therefore, the maximum change in flux for a given latitude would happen between the periastron and the apastron: 

\begin{equation}\label{eq:deltaF}
    \Delta F=\frac{L_{\star}}{4\pi a^{2}}\frac{e}{1-e^{2}}\cos|~\beta~| 
\end{equation}

Considering the nominal value of the eccentricity and semi-major axis given in Table~\ref{table:sys_prop}, we found that \gjb\ receives at the equator 2088--$1051\,\mathrm{Wm}^{-2}$ (1.53--0.77~F$_{\oplus}$), 1476--$743\,\mathrm{Wm}^{-2}$ (1.08--0.54~F$_{\oplus}$) at $\beta=\pm45\degr$, and $\sim0\,\mathrm{Wm}^{-2}$ (0.0~F$_{\oplus}$) at the poles. These values suggest that the stellar flux received by the planet may vary by as much as 50\% along its orbit, which might have strong implications in terms of the global climate of the planet. 
 
Furthermore, in Section~4 we obtained the tidal heating experienced by the innermost planets in the system. The effect of tidal heating in rocky or terrestrial planets and exo-satellites may have significant implications for their habitability, notably for M dwarfs \citep{Shoji2014Thermal-orbitalStars}, for whom the HZ and tide-regions overlap. Thus, tidal heating could either favor or work against the habitability of \gjb. As an example, and making an analogy with the Jupiter-moons system, Europa is a rocky body covered by a 150-km thick water-ice crust, tidal heating may maintain a subsurface water ocean. In the case of the Jovian satellite Io, the extreme violence of the tides provokes intense global volcanism and rapid resurfacing, ruling out any possibility of habitability. Indeed, in the case of Io, the tidal heating was found to be $h=2\,\mathrm{Wm}^{-2}$ and it is considered that larger heating rates would result in uninhabitable planets \citep{Blaney1995VolcanicComposition, Bagenal2004JupiterMagnetosphere}. On the other hand, internal heating is used to explain Earth's plate tectonics, which may actually enable planetary habitability since it drives the carbon--silicate cycle, which stabilizes the temperature of the atmosphere and the levels of CO$_{2}$. Indeed, from Martian geophysics studies it has been suggested that tectonic activity ceased when the internal heat dropped below $0.04\,\mathrm{Wm}^{-2}$ \citep{Williams1997HabitablePlanets}. Based on these arguments, \cite{Barnes2009TidalHabitability} established the limits of tidal heating which may be compatible with a biosphere in the range of 0.04-$2.0\,\mathrm{Wm}^{-2}$. In this context, we found that \gjb, with tidal heating of $h=0.1\,\mathrm{Wm}^{-2}$, is compatible with the development of life. 

In Section~5 we highlighted the importance of minor bodies on the emergence and maintenance of life on a planet, and we identified stable locations in the system where minor bodies may reside after the planetary formation. It is beyond the scope of this paper to explore deeper how the potential hosted bodies of the stable regions evolve and behave due to their gravitational interactions with other planets in the system, such as those that occur in the Solar System. But this study would be highly interesting due to the location of \gjb\ in between two minor-body reservoirs. An example of this are the theoretical studies of three systems HD\,10180, 47\,UMa, and HD\,141399 by \cite{Dvorak2020OnSystems}, where the authors studied the dynamical evolution of hypothetical exocomet populations, and the study presented by \cite{Schwarz2017ExocometsTransport}, where the authors explored the water delivery to the planet Proxima Centauri-b in consideration of a hypothetical population of exocomets. A similar study was performed by \cite{Dencs2019WaterPlanets}, where the authors investigated a synthetic late-heavy-bombardment-like event in the TRAPPIST-1 system, which could have transported a  significant amount of water to the habitable planets discovered in the system. All these studies could be combined with the recent prescriptions given by \cite{Wyatt2019SusceptibilityShoreline} to perform predictions about the evolution of \gjb's atmosphere due to planetesimal impacts.  

Another important factor to consider in terms of habitability are stellar flares and coronal mass ejections (CMEs), where the former deliver strong and isotropic ultraviolet (UV) radiation, and the latter form directional streams of charged particles. While a potential atmosphere on \gjb\ could protect the surface from harmful UV light, flares pose an imminent danger to surface life if these absorbers are depleted. For example, \citet{Tilley2019ModelingPlanet} found that the CMEs associated with strong flares can deplete the ozone in a potential atmosphere, leaving the flares' UV radiation to hit the surface nearly unabsorbed. On the other hand, prebiotic chemistry research has shown that a certain amount of flaring might, in fact, be necessary to form possible precursors for life in the first place \citep{Rimmer2018TheExoplanets}. This suggests that surface life might rely on a host stars with the right balance of flaring, not too much and not too little \citep[see e.g.][]{Gunther2020Dwarfs}.

\section{Conclusions}

In this study, we sought to better understand the real nature of the planetary system \gj. This system is of special interest because it is the fourth closest planetary system to the Sun orbiting an M dwarf (3.75~pc) known to host a planet in the HZ, just after Proxima Centaury  \citep[1.30~pc, ][]{Anglada-Escude2016ACentauri}, Ross-128 \citep[3.37~pc, ][]{Bonfils2018AParsec}, and GJ~1061 \citep[3.67~pc, ][]{Dreizler2020RedGJ1061}. The planets in \gj\ have been detected so far only via RVs  (\gjb\ and \gjc), and there are two candidates, \gjd\ and \gje. We performed a comprehensive study of \gj's planetary formation and water delivery history in its early times, a dynamical stability of the system and the effects of tides on the planets, as well as the dynamical environment, where we identified regions where the system may harbour minor bodies in stable orbits, i.e., MAB and KB analogues. Each of these parameters has an important influence on the potential habitability of the planet located in the HZ, \gjb. We also reviewed the published observational data obtained via \harps\ and \tess, where we explored the detectability of the known, candidates and extra planets in the system. More specifically, our studies revealed:  

   \begin{enumerate}
      \item We analyzed the process of formation and evolution of the planetary system around \gj\ making use of N-body simulations. Our research allowed us to derive three important results: (1) the gaseous component of the disk played a key role in determining the physical and orbital properties of the planets of the system; (2) \gjc\ is a planet with very low water content, while \gjb\ seems to be a water world evolving within the HZ; and (3) the planetary system around \gj\ might host more planets (in addition to the two confirmed) with different physical and orbital properties.
      
      \item From our review of the observational data sets we found that these other possible planets predicted by the N-body simulations could not have been detected with the precision and cadence of the HARPS RV data-set that confirmed \gjb\ and \gjc. From the \tess\ data, however, we found that a transit of the innermost planet \gjc\ can be ruled out. On the other hand, for the rest of the planets in the system, the \tess\ photometric data were insufficient to confidently rule out the presence of any transit signals. 
      
      \item We tested the stability of the system using the MEGNO criterion. We found that in the two-planet configuration we could not constrain the inclination of the system, and consequently the masses of the planets remain unknown. However, in the four-planet configuration, the system was stable only for near-circular orbits of the two outermost planets, \gjd\ and \gje, and minimum inclinations of the whole system down to $\sim$72$\degr$. Hence, this scenario allowed us to unveil the system as likely composed by an Earth-mass planet, a super-Earth, and two mini-Neptunes with masses of: $1.18\leq \mc\leq1.24\,\mearth$, $2.89\leq \mb\leq3.03\,\mearth$, $10.80\leq \md\leq11.35\,\mearth$, and $9.30\leq \me\leq9.70\,\mearth$, respectively. In this scenario, using M--R relationships for different bulk compositions we found that the radii of the planets would be in the following ranges: $1.04\leq \rc\leq1.18\,\rearth$, $1.32\leq \rb\leq1.83\,\rearth$, $2.00\leq \rd\leq3.68\,\rearth$, and $1.91\leq \re\leq3.35\,\rearth$, respectively. 
      
      \item Since the two innermost planets occupy close-in orbits, we performed a tidal-evolution analysis where we found that both \gjb\ and \gjc\ are likely in a pseudo-rotational state, i.e., their spins are aligned with the host star's and they have reached constant rotational periods. This state was reached over a short time scale of $<10^{7}\,\yr$, which is much shorter than the age of the star, which we estimated to be older than a few Gyr. However, we found that the time required to circularize the orbits was much longer, which could last several $10^8~\yr$ or $10^9\,\yr$, which hints that the orbits might not be fully circular, 
      and hence a certain amount of tidal heating may be induced.  

      \item Using the MEGNO criterion we studied the similarities of \gj\ with the Solar System in terms of minor-body reservoirs in both the two- and the four-planet configurations. We found several regions where minor bodies might reside in stable orbits, forming structures like the MAB or KB. Each structure was sculpted differently in each scenario and, in particular, there were two structures present in all simulations: 1) region B, which was trapped in between planets \gjb\ and \gjc, and may be considered as an hot-MBA analogue, and 2) region C, which may be a hot-KB analogue in the two-planet configuration, or an hot-MBA analogue trapped between \gjb\ and \gjd\ in the four-planet configuration. All these structures identified in \gj\ seem to have some similarities with the ones that we have in the Solar System, but at the same time they point out the large diversity of minor-body reservoirs which may exist.

   \end{enumerate}     

Finally, taking together all the analyses performed in this study and if the four-planet configuration is eventually confirmed, we conclude that \gjb\ is likely a super-Earth residing in a stable orbit close to the inner border of the HZ of its host star. The planet might be hydrated and likely in a pseudo-rotational state with in a non-circular orbit. Such a small eccentricity could provoke remarkable variations of the stellar flux received that could result in strong implications in terms of the global climate of the planet. Nevertheless, we also found that \gjb\ undergoes tidal heating compatible with the development and existence of a biosphere. The reservoirs of minor bodies could also play a role in the emergence and maintenance of life in \gjb. In fact, it would be desirable to investigate further how minor bodies such as asteroids and comets evolve and behave with \gjb, in particular the impact-flux ratio, where aspects such as the watering, the delivery of organics and the evolution of planetary atmosphere  take place.


\begin{acknowledgements}
We thank the anonymous referee for the helpful comments. J.C.S. acknowledges funding support from Spanish public funds for research under projects ESP2017-87676-2-2 and RYC-2012-09913 (Ramón y Cajal programme) of the Spanish Ministry of Science and Education. Z.M.B. acknowledges CONICYT- FONDECYT/Chile Postdoctorado 3180405. M.G. and E.J. are FNRS Senior Research Associates. V.V.G. is a FNRS Research Associate. M.N.G. acknowledges support from MIT’s Kavli Institute as a Torres postdoctoral fellow. We acknowledge the use of public data from pipelines at the TESS Science Office and at the TESS Science Processing Operations Center. 
\end{acknowledgements}


\bibliographystyle{aa}
\interlinepenalty=10000
\bibliography{my_references}

\begin{appendix}
\section{Location of the snow line}
\label{ap:sl}
In our modelling, we followed the procedure proposed by \citet{Ciesla2015VolatileStars} to compute the location of the snow line, which sets the boundary between dry embryos and water-rich embryos in each of our N-body simulations. This procedure is described in detail below.

To do this, we assumed that water condenses in a protoplanetary disk, where its partial pressure exceeds the saturation pressure, which is associated with temperatures between 140~K and 180~K for typical disk conditions. In our work, we adopt a value of $T$ = 140~K to define the position of the snow line in the protoplanetary disk. To determine the distance $R$ from the central star in the disk mid-plane, where $T$ = 140~K, we adopt the temperature profile suggested by \citet{Chiang1997SpectralDisks}, which is given by:
\begin{equation}
T(R) = \left(\frac{\alpha}{4}\right)^{1/4}\left(\frac{R_{\star}}{R}\right)^{1/2}T_{\star}, 
\end{equation}
where $R$ is the radial coordinate in the disk mid-plane, $R_{\star}$ and $T_{\star}$ are the radius and effective temperature of the star, respectively, and $\alpha$ is the grazing angle at which the starlight strikes the disk. According to this, the location of the snow line ($R_{\text{ice}}$) will be given by:
\begin{equation}
  R_{\text{ice}} = \left(\frac{\alpha}{4}\right)^{1/2} R_{\star} \left(\frac{T_{\star}}{140 ~\text{K}}\right)^{2}.
  \label{Ec_Rice}
\end{equation}

\begin{figure}
\includegraphics[width=\columnwidth]{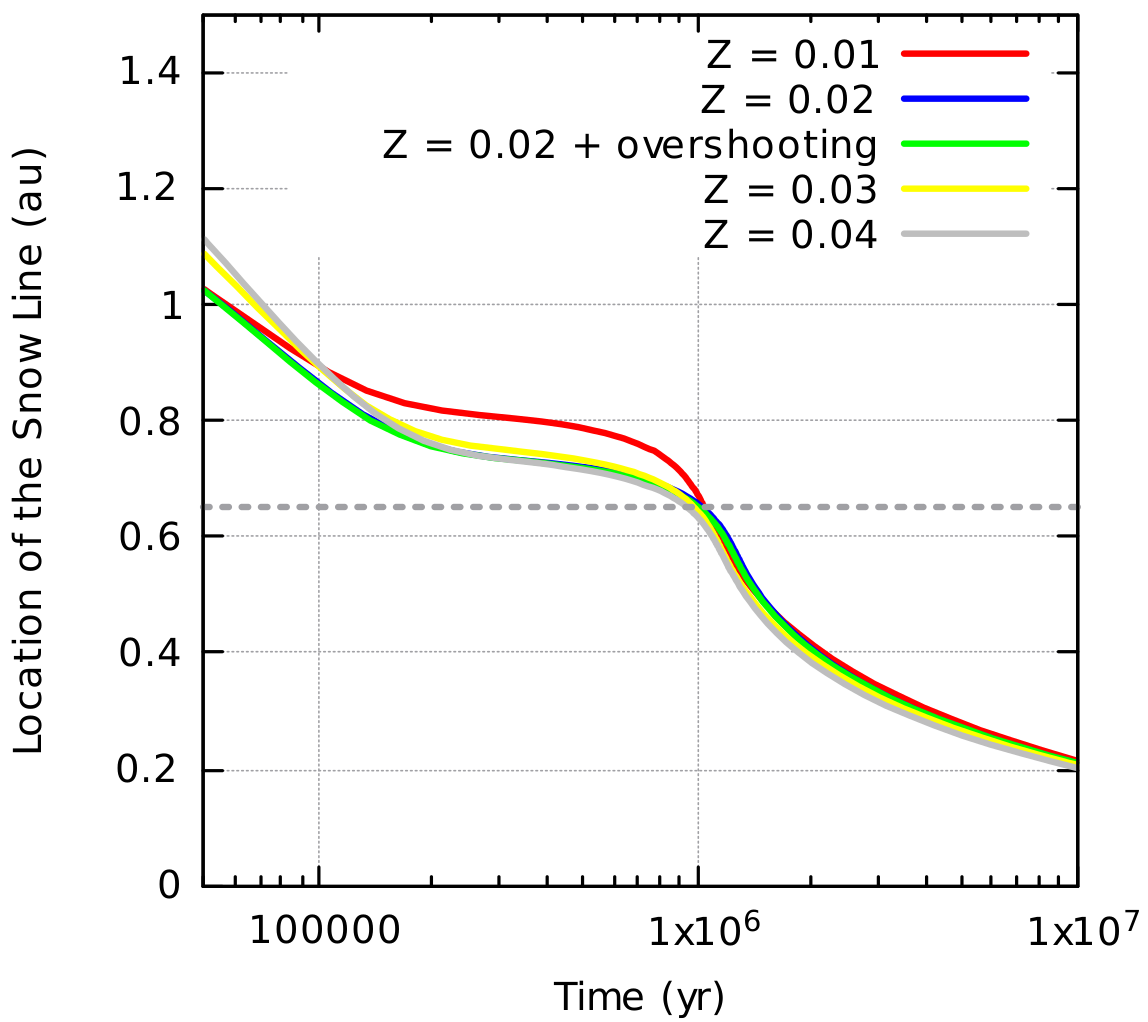}
\caption{Location of the snow line as a function of time for a 0.3 M$_{\odot}$ star based on the stellar models developed by \citet{Siess2000AnStars}. The five curves correspond to different values of the star initial metallicity, $Z$, (in terms of mass fraction). The gray dashed line indicates a value of 0.65~au.
}
\label{SnowLine}
\end{figure}

As the reader can see, $R_{\text{ice}}$ strongly depends on the radius and the effective temperature of the central star, which evolve over time. Following the stellar evolutionary models developed by \citet{Siess2000AnStars}, we made use of Eq.~\ref{Ec_Rice} to determine the temporal evolution of the snow line in a protoplanetary disk around a 0.3 M$_{\odot}$ star, which is illustrated in Fig.~\ref{SnowLine}. The five curves presented in the figure correspond to different values of the star's initial metallicity, $Z$, (in terms of mass fraction) adopted in the stellar models of \citet{Siess2000AnStars}. It is important to note that we assume a value of $\alpha$ = 0.045 in the construction of these five curves, which is in agreement with the expressions given by \citet{Chiang1997SpectralDisks}, and with the value used by \citet{Ciesla2015VolatileStars}. 

Like \citet{Ciesla2015VolatileStars}, we set the location of the snow line at 1 Myr. We consider that this choice can be somewhat arbitrary since the snow line will continually migrate inward as the system evolves. However, our selection simply defines a boundary between the dry embryos and water-rich embryos in our simulated protoplanetary disk. According to the results illustrated in Fig.~\ref{SnowLine}, the snow line at 1 Myr is located at 0.65~au for all values of the star's initial metallicity, $Z$, adopted by \citet{Siess2000AnStars}. Table~\ref{tab:Table_Siess} shows the values of the luminosity, temperature, and radius of a 0.3~M$_{\odot}$ star at 1 Myr used in our work to compute the location of the snow line, which were extracted from the different stellar models of \citet{Siess2000AnStars}.

\begin{table}
\centering
\caption{Luminosity, temperature, and radius at 1 Myr of a 0.3~M$_{\odot}$ star for different values of the star initial metallicity (in terms of mass fraction) \citep{Siess2000AnStars}}.
\label{tab:Table_Siess}
\begin{tabular}{|c|c|c|c|} 
\hline  
Metallicity & Luminosity & Temperature & Radius \\
     & (L$_{\odot}$) & (K) & R$_{\odot}$\\
\hline
0.01 & 0.722040 & 3604 & 2.061  \\
0.02 & 0.682419 & 3360 & 2.305  \\
0.02 + OS & 0.678208 & 3359 & 2.298 \\
0.03 & 0.660323 & 3365 & 2.259 \\
0.04 & 0.636591 & 3289 & 2.329  \\
\hline
\end{tabular}
\end{table}

From Fig.~\ref{SnowLine}, it is evident that the snow line changes over time. However, we follow \citet{Ciesla2015VolatileStars} and other previous studies based on N-body simulations, and assign a single location for the snow line ($R_{\text{ice}}$ = 0.65~au) for the whole duration of the numerical experiments. As we previously mentioned, the snow line is defined to determine a boundary between dry embryos and water-rich embryos in the protoplanetary disk. 

\section{Detrended models}

\begin{figure*}
  \centering 
  \includegraphics[width=0.95\textwidth]{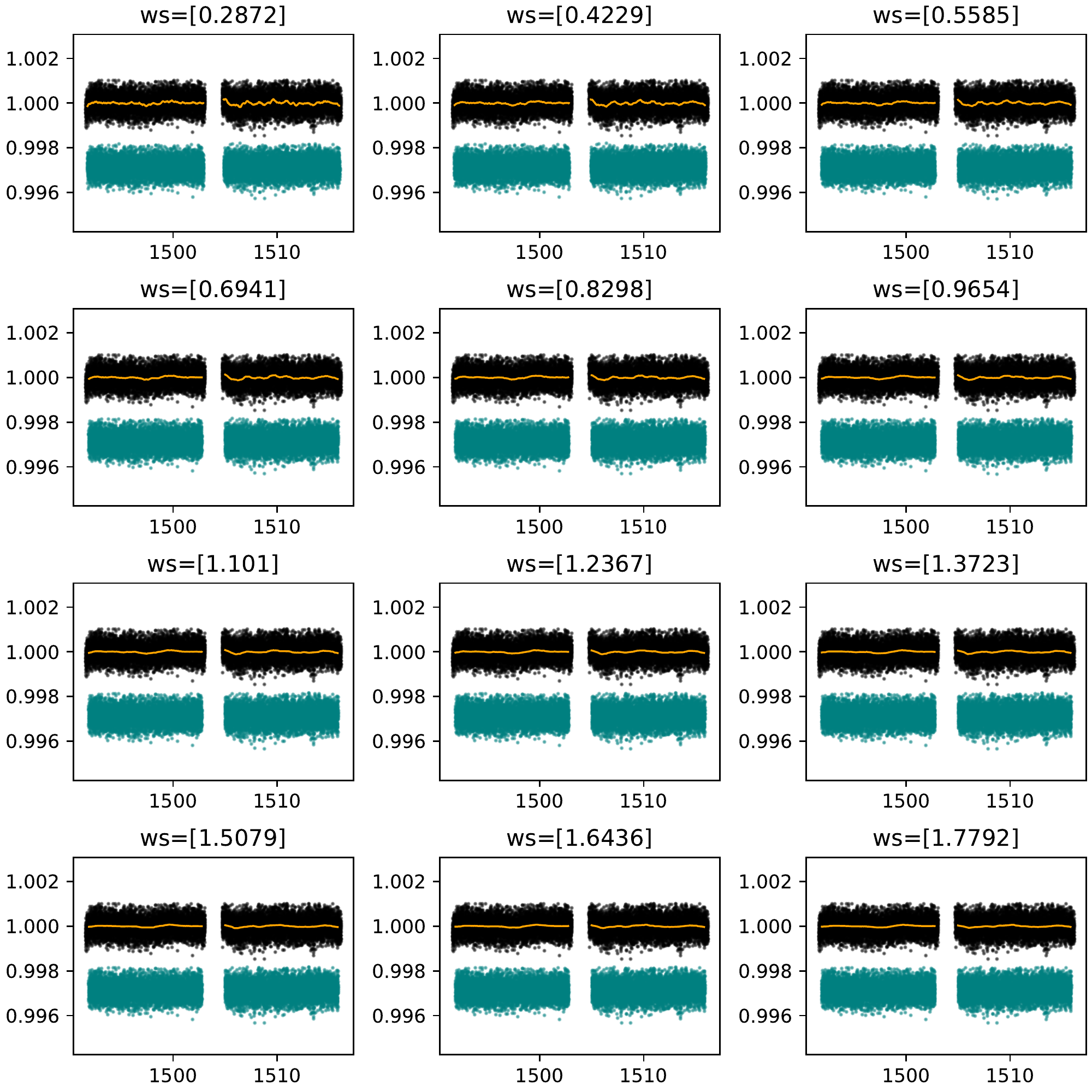}
  \caption{Detrended models applied to the TESS data in our search for threshold-crossing events. We used the bi-weight method with different window-sizes. Each panel indicates the window-size (in units of d) in the top. The black points show the PDC-SAP fluxes, the solid-orange line are the identified trends, and the teal points are the detrended data used as input into the \tls. In all cases, the $y$-axis shows the normalized flux, and the $x$-axis is units of Barycentric TESS Julian Date.}\label{detrend}
\end{figure*}

\end{appendix}

\end{document}